\documentclass[lettersize,journal]{IEEEtran}
\usepackage{amsmath,amsfonts}
\usepackage{algorithmic}
\usepackage{algorithm}
\usepackage{array}
\usepackage{hyperref}
\usepackage[caption=false,font=normalsize,labelfont=sf,textfont=sf]{subfig}
\usepackage{textcomp}
\usepackage{stfloats}
\usepackage{url}
\usepackage{verbatim}
\usepackage{multirow}
\usepackage{graphicx}
\usepackage{cite}
\usepackage{times}                     

\usepackage{tabu}                      
\usepackage{booktabs}                  
\usepackage{lipsum}                    
\usepackage{mwe}                       
\usepackage{subcaption}

\usepackage{mathptmx}                  
\usepackage{amsmath}
\usepackage{amssymb}
\title{Vision6D: 3D-to-2D Interactive Visualization and Annotation Tool for 6D Pose Estimation}

\author{Yike Zhang, Eduardo Davalos, Jack Noble,~\IEEEmembership{Senior Member,~IEEE}
\thanks{This paper was produced by the IEEE Publication Technology Group. They are in Piscataway, NJ.}
\thanks{Manuscript received April 19, 2021; revised August 16, 2021.}}

\begin{document}



\maketitle

\begin{abstract}
Accurate 6D pose estimation has gained more attention over the years for robotics-assisted tasks that require precise interaction with physical objects. This paper presents an interactive 3D-to-2D visualization and annotation tool to support the 6D pose estimation research community. To the best of our knowledge, the proposed work is the first tool that allows users to visualize and manipulate 3D objects interactively on a 2D real-world scene, along with a comprehensive user study. This system supports robust 6D camera pose annotation by providing both visual cues and spatial relationships to determine object position and orientation in various environments. The annotation feature in Vision6D is particularly helpful in scenarios where the transformation matrix between the camera and world objects is unknown, as it enables accurate annotation of these objects' poses using only the camera intrinsic matrix. This capability serves as a foundational step in developing and training advanced pose estimation models across various domains. We evaluate Vision6D's effectiveness by utilizing widely-used open-source pose estimation datasets Linemod and HANDAL through comparisons between the default ground-truth camera poses with manual annotations. A user study was performed to show that Vision6D generates accurate pose annotations via visual cues in an intuitive 3D user interface. This approach aims to bridge the gap between 2D scene projections and 3D scenes, offering an effective way for researchers and developers to solve 6D pose annotation related problems. The software is open-source and publicly available at \href{https://github.com/InteractiveGL/vision6D}{https://github.com/InteractiveGL/vision6D}.
\end{abstract}

\begin{IEEEkeywords}
3D-to-2D Registration, 6D Pose Estimation, Pose Annotation, Computer Vision, Human-Computer Interaction, Augmented Reality.
\end{IEEEkeywords}

\section{Introduction} 
The estimation of 6D poses, which involves determining the position and orientation of an object in a 3D space, is a fundamental task in computer vision, particularly for applications in robotic-assisted systems, augmented reality, autonomous navigation, and 3D scene understanding \cite{s24041076, wang2019normalizedobjectcoordinatespace}. Accurate pose estimation assists vision-guided robotics to interact with their physical environment by identifying and manipulating objects. However, this task is inherently challenging due to factors such as occlusion, object symmetries, varying lighting conditions, and cluttered scenes, which complicate the retrieval of accurate object poses. Traditionally, the problem of 6D pose estimation has been approached by establishing 2D-to-3D correspondences and performing iterative Point-n-Perspective (PnP) algorithms to estimate a pose \cite{xiang2018posecnnconvolutionalneuralnetwork, Collet2011, Li_2019}. However, such methods require textured objects to detect matching feature points in the 2D images. 

The proposed annotation tool aims to overcome this shortage and offers manual registration between texture-less 3D objects and 2D images. Recent advancements in deep learning and computer vision have led to substantial improvements and promising results in 6D pose estimation tasks\cite{9836663, wang2019densefusion6dobjectpose, peng2018pvnetpixelwisevotingnetwork}. However, many existing approaches rely heavily on ground-truth pose information, which is often unavailable when applied to custom datasets or real-world scenarios, such as identifying target structures in prerecorded videos. In these scenarios, obtaining ground-truth camera poses is nontrivial because prerecorded videos typically provide only 2D images, 3D object models, and at times, the camera's intrinsic parameters. Consequently, achieving accurate 3D-to-2D registration in the video's first frame is critical for subsequent frames' object poses. These limitations have inspired us to develop a tool that assist the interactive visualization and annotation of 6D object poses, supporting various downsteaming tasks, such as training artificial intelligence (AI) models for 6D pose estimation related tasks.

In this paper, we propose an interactive 3D-to-2D visualization and annotation tool specifically designed for 6D pose estimation tasks named Vision6D. The tool provides a user-friendly 3D interface that supports 3D-to-2D registration by utilizing the image formation model \cite{Peng2009}, allowing users to visualize, annotate, and refine 3D object poses interactively. When a camera intrinsic matrix is provided or estimated, the tool assists users in reconstructing the original camera pose by enabling annotations directly in 3D space and aligning these annotations with the corresponding 2D image, thus producing accurate 6D pose annotations.
Our proposed system presents four main \textbf{contributions} as follows:
\begin{enumerate}
    \item Vision6D provides an interactive framework that effectively aligns 3D models onto 2D images, enabling precise 6D pose annotation. This bridges the gap between 2D image projection and the spatial complexity of 3D scenes.
    \item The tool allows users to efficiently annotate and refine 6D poses via an interactive user interface, simplifying the dataset generation process. To the best of our knowledge, Vision6D is the first tool designed specifically for this purpose.
    \item We validate the effectiveness of Vision6D through a comprehensive user study, demonstrating that it offers an intuitive and accurate solution for 6D pose annotation.
    \item The user study used public 6D pose estimation datasets named Linemod \cite{linemod} and HANDAL \cite{guo2023handaldatasetrealworldmanipulable}, where user-annotated poses were compared against ground-truth poses. The results illustrate the tool's accuracy, efficiency, and usability, highlighting its potential as a standardized solution for 6D pose annotation.
\end{enumerate}
\section{Related Work}
\subsection{3D Graphical User Interface}
3D graphical user interfaces (GUIs) allow users to interact with and understand complex 3D scenes. These GUIs aim to provide users with intuitive controls and visualization to handle and process 3D data, making it easier to perform tasks across different domains, such as animation, video game development, robotics, and computer vision related tasks. Efficiently navigating and interacting with 3D scenes has been heavily used for both technical and creative applications. Modern 3D GUIs also provide many functionalities that improve user experience.

One of the primary uses of 3D GUIs is to help users gain a better understanding of 3D scenes. By offering interactive visualizations of 3D objects and spatial environments, these GUIs enable users to more accurately analyze spatial relationships, identify object positions, and determine camera viewpoints. For example, Blender \cite{blender, Blender4Robotics}, a popular open-source 3D creation application, provides an extensive toolbox for modeling, texturing, and animating 3D scenes. The GUI helps users manipulate 3D models under different camera view angles, effectively controlling animations and other visual effects. The ability to control lighting, shading, and smooth object transformations makes Blender an essential computer graphics software for 3D animations and rendering.
Other than Blender, there is another popular platform for developing interactive 3D scenes called Unity \cite{unity, unity3d_model}. It provides extensive controls for camera manipulation, object transformations, and real-time rendering. Unity’s 3D GUI allows users to interactively design 3D scenes, set camera trajectories, and simulate real-world physics, which are essential for creating immersive virtual experiences. The platform is ideal for game development, providing versatility and user-friendly features.

In robotics and computer vision related fields, 3D GUIs play an important role in 3D scene understanding, such as camera poses. Pangolin is a lightweight 3D visualization library for real-time display of 3D objects, camera trajectories, and keypoints \cite{pangolin, Khole_2023}. The library is broadly used in conjunction with Simultaneous Localization and Mapping (SLAM) systems to display 3D camera pose trajectories directly. The GUI of Pangolin allows users to track camera movements and inspect 3D point clouds in real-time, making it useful for applications such as autonomous navigation and mapping systems. The simplicity and fast speed of Pangolin make it ideal for technical tasks where real-time feedback is preferred to understand object localization in space.

Overall, 3D GUIs aim to support users by providing flexible and interactive environments for performing simulated 3D-related tasks. Whether creating animations in Blender, developing games in Unity, or tracking and visualizing camera poses in Pangolin, these interfaces provide a way to understand and interact with 3D data. By offering user-friendly visualizations, configurable controls, and convenient interactions, 3D GUI-based applications make it easier for users to create and explore 3D scenes, enabling more efficient workflows across various domains. These 3D GUI software excels within their respective domains but does not explicitly address the need for efficient 6D pose annotation in 3D-to-2D registration tasks. The lack of interactive features specifically tailored for 6D pose annotation restricts users' ability to quickly obtain and modify camera poses.

\subsection{6D Pose Annotation Techniques}
6D pose estimation involves determining the 3D rotation and 3D translation of an object or the corresponding camera pose in space. Accurate 6D pose estimation allows vision-based robotics to interact with objects in the real world. However, generating reliable 6D pose annotations in cluttered scenes and achieving precise estimation remain challenging. Traditionally, 6D pose annotations have been labor-intensive, sometimes relying on manual trial and error \cite{hodan2017tlessrgbddataset6d, 8265467}. In such methods, users manually adjust the 6D pose parameters by modifying numerical values for position and orientation, then render the scene to verify accuracy. This iterative process is time-consuming and error-prone, especially for objects in cluttered or occluded environments, where small adjustments are difficult to determine visually. While such manual approaches can produce accurate results with patience, they are not optimal for fast development. Another common approach to 6D pose estimation and annotation has involved using ArUco markers \cite{GARRIDOJURADO20142280} or similar fiducial marker systems to determine camera poses in recent years. ArUco boards consist of grids of unique markers. Their known geometric properties are used to calculate the camera’s extrinsic parameters. While this method can yield accurate camera poses in controlled environments, placing these markers prior to capturing video might not be a possibility, such as in surgical, automotive, and aerospace domains. Moreover, this method needs the markers to be consistently visible to the camera, which is often infeasible in cluttered and uncontrollable scenarios.

Annotating the 6D pose in a video's first frame brings additional challenges. Traditionally, methods for first-frame pose estimation require either manual initialization or external equipment, such as a checkerboard for camera calibration. The accuracy of first-frame estimation plays a vital role in the subsequent frames' estimated camera poses, as errors can accumulate across sequential frames, lowering overall performance. Methods that rely solely on image data for the first frame, without external markers or camera calibration, often struggle in complex environments where occlusions or ambiguities are present. One of the notable challenges in 6D pose annotation is the lack of real-time feedback and support for exporting camera poses in many existing annotation tools. While some advanced systems offer sophisticated modeling and rendering capabilities, they often fail to provide real-time visualization of pose manipulation or to easily obtain camera pose data for various downstream tasks. This limitation can cause problems in both manual and semi-automated annotations, as users are forced to repeatedly modify and render scenes to verify pose accuracy, slowing down the overall workflow.

In summary, while numerous methods exist for generating 6D camera pose estimations, many have inherent limitations. Interactive pose visualization and seamless pose updates remain not supported in many existing tools, particularly for 3D-to-2D registration. These challenges inspired us to focus on developing interactive tools that can provide more efficient and accurate 6D pose annotations.

\subsection{Addressing the Literature Gap}
Despite recent advances in deep learning for 6D pose estimation, existing annotation methods, often reliant on fiducial markers or manual frame-by-frame rendering, remain inefficient and impractical for real-world scenarios like surgical scenes. ArUco markers work well in controlled settings but are unsuitable for domains where marker placement is infeasible, and manual pose adjustment is time-consuming and lacks interactivity. These limitations constrain pose annotation and make current workflows inadequate for applications requiring immediate feedback or large-scale dataset generation.

We propose a 3D-to-2D interactive visualization and annotation tool for 6D pose estimation. This tool identifies and addresses these gaps above by allowing users to visualize 2D image data and 3D models synchronously. The tool provides a more intuitive environment for generating and refining pose annotations. Users can interactively adjust poses and receive immediate feedback, helping the process of annotating frames in video sequences. Our tool's ability to create first-frame or frame-by-frame pose annotation substantially improves over manual trial-and-error rendering or marker-based methods. Moreover, it eliminates the need for external markers in scenarios where they are impractical. In general, while traditional 6D pose annotation methods and existing 3D GUIs offer valuable features, they do not meet the growing demand for an efficient and interactive tool in complex, real-world settings. The proposed tool fulfills this gap by providing an innovative solution for annotating 6D poses via an application that is more user-friendly and applicable across different domains.

\section{Vision6D}
Vision6D is a 3D-to-2D annotation software that provides interactive visualization and annotation for the 6D camera pose estimation research community. Firstly, we will discuss the underlying foundational mathematical framework that the software builds on, including image formation, 3D projection, and coordinate transformations that represent the 3D-to-2D problem space and develop a 6D camera pose. Secondly, we will present the technical details related to the system's design and implementation.

\subsection{Mathematical Formalism}
In this subsection, we focus on the mathematical formalism behind Vision6D, and mainly discuss image formation, camera intrinsic and extrinsic matrices, projection and rendering, to establish visible 3D-to-2D correspondences for the 6D camera pose estimation.
Image formation is the process of projecting a 3D scene onto a 2D image plane through the lens of a camera. In this context, the camera's intrinsic (e.g., focal length) and extrinsic (e.g., orientation) parameters define the relationship between the 3D world and the 2D image representation. The camera intrinsics are the internal characteristics of the camera, including the focal length $f_x$ and $f_y$, principal point $c_x$ and $c_y$. These parameters are combined and represented by a 4 $\times$ 4 matrix:
\begin{equation}
    K = \begin{bmatrix}
        f_x & 0 & c_x & 0 \\
        0 & f_y & c_y & 0 \\
        0 & 0 & 1 & 0 \\
        0 & 0 & 0 & 1
    \end{bmatrix},
\end{equation} which transforms a 3D point from camera coordinates into the image coordinates system. Camera extrinsic parameters describe the orientation and position of the camera in the world coordinates, showing how the camera is placed relative to the scene. The transformation matrix $M$ combines a 3D rotation matrix:
\begin{equation}
    R = \begin{bmatrix}
        r_1 & r_2 & r_3\\
        r_4 & r_5 & r_6\\
        r_7 & r_8 & r_9
    \end{bmatrix},
\end{equation}
and a 3D translation vector $t = [t_1, t_2, t_3]$ to form a 4 $\times$ 4 matrix $M=[R|t]$ that maps points from the world coordinates to the camera coordinates system.
More precisely, $M$ transforms a homogeneous 3D point from world coordinates $P_w$ into camera coordinates $P_c$ via $P_c = MP_w$. To transform a point from camera to world coordinates system, the inverse transformation matrix $M^{-1}$ would be used, thus $P_w = M^{-1}P_c$.
\begin{figure}[ht]
    \centering
    \includegraphics[width=\linewidth]{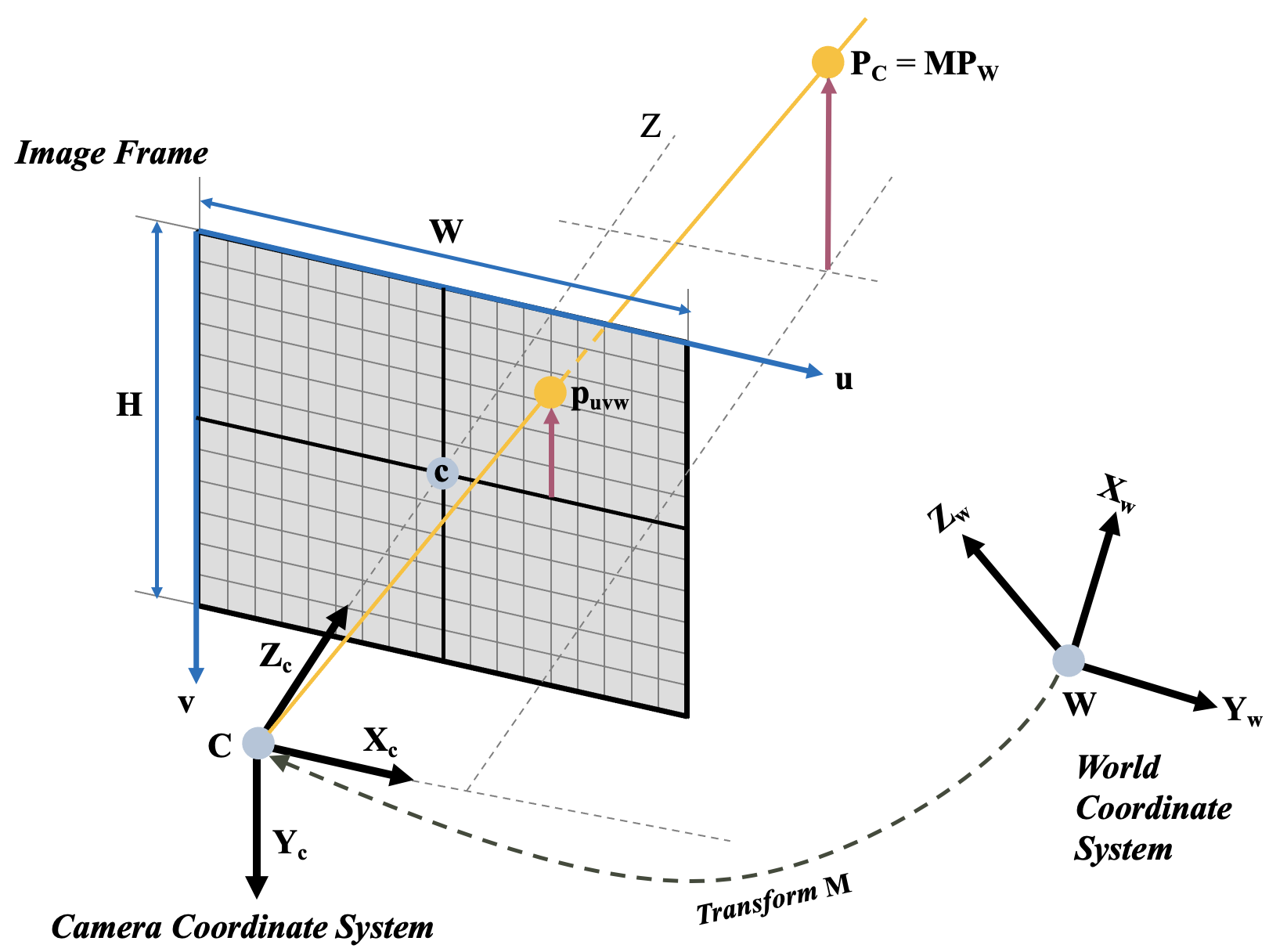}
    \caption{\textbf{3D-to-2D Projection.} Demonstration of projecting 3D object point onto a 2D image plane using the camera intrinsic and extrinsic parameters.}
    \label{fig:projection}
\end{figure}

After the 3D scene is transformed into the camera's coordinate system using the transformations above, the next key step is projection and rendering. These are the processes for mapping 3D points $\left[X, Y, Z\right]$ from the camera coordinate system onto the 2D image plane. A pinhole camera model is used to project the 3D camera coordinates onto the image plane. The projection from 3D-to-2D is mathematically defined as $P_{uvw} = KP_c$ where $P_{uvw}$ is the homogeneous 2D pixel coordinate of the projected point on the $uv$ image plane. 
\begin{equation}
    P_{uvw} = \begin{bmatrix}
        f_x & 0 & c_x & 0\\
        0 & f_y & c_y & 0\\
        0 & 0 & 1 & 0\\
        0 & 0 & 0 & 1\\
        \end{bmatrix}
        \begin{bmatrix}
        r_1 & r_2 & r_3 & t_1\\
        r_4 & r_5 & r_6 & t_2\\
        r_7 & r_8 & r_9 & t_3\\
        0 & 0 & 0 & 1\\
        \end{bmatrix}
        \begin{bmatrix}
        X\\
        Y\\
        Z\\
        1\\
        \end{bmatrix}
\end{equation}
This projection process involves applying the intrinsic parameters of the camera $K$ to $P_{c}$ to map the point to the 2D image as shown in Figure~\ref{fig:projection}. The rendering mechanism ensures that 3D models are correctly displayed in the 2D view according to the camera's perspective.

\subsection{User Interface and Layout}
\begin{figure*}[!htbp]
    \centering
    \includegraphics[width=0.8\linewidth]{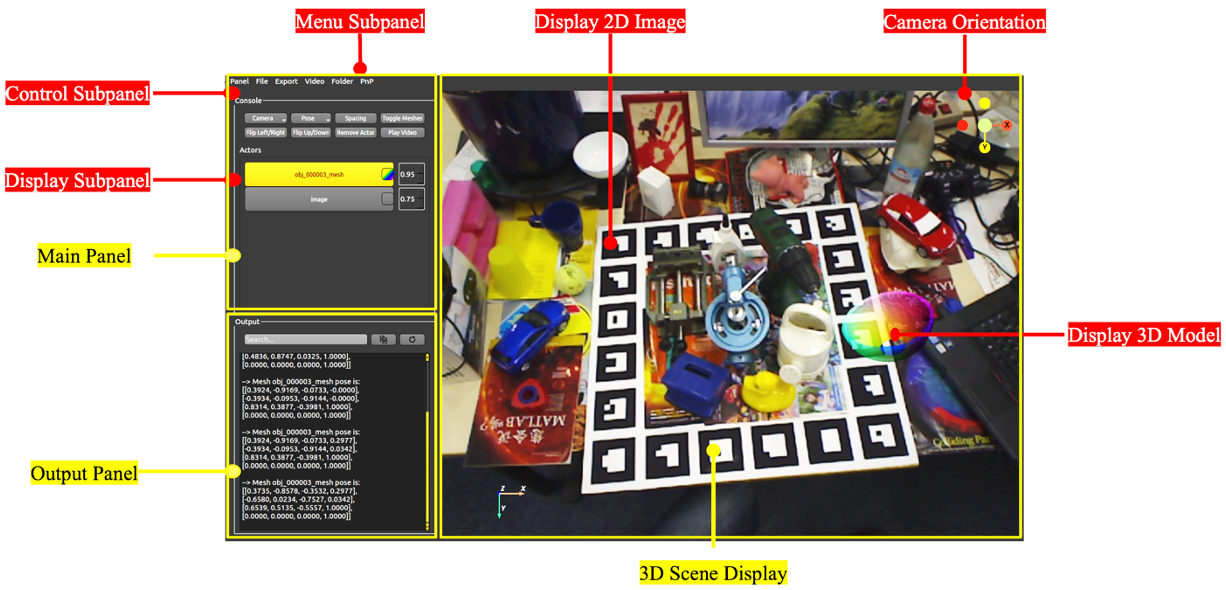}
    \caption{\textbf{Vision6D's 3D User Interface}. Screenshot and decomposition of essential 2D and 3D features to support the interactive 6D pose annotation workflow.}
    \label{fig:system_overview}
\end{figure*}

A screenshot and a decomposition of the Vision6D user interface are presented in Figure~\ref{fig:system_overview}. The user interface is composed of three key components to simplify the process of 6D pose annotation: the Main Panel, the 3D Scene Display, and the Output Panel. In this subsection, we will use the Linemod-Occluded dataset \cite{10.1007/978-3-319-10605-2_35} as an example to illustrate the functionality and features of the Vision6D application.
\subsubsection{Main Panel}
As shown in Figure~\ref{fig:system_overview}, the Main Panel consists of three components: the Menu, Control, and Display Sub-panels.
The Menu Subpanel provides an interface for input and output operations, such as loading and saving renderings/3D meshes and exporting pose annotations. Users can easily access menus to import new data and save workspaces. The Menu Subpanel offers quick access to common file management features. The saved data can be used for further analysis or to integrate into deep learning training pipelines.

The Control Subpanel allows users to apply various modifications to the 3D scene. This includes options for adjusting camera intrinsics and managing other settings, such as defining poses, toggling visibility of meshes, mirroring meshes, and adjusting 3D models' spacing. For example, users are able to toggle the visibility of 3D objects or update other parameters on-the-fly to visualize the effects in the 3D Scene Display. This Control Subpanel gives users the freedom to adjust the voxel spacing in 3D objects or flip objects' orientations as needed. These controls provide a hands-on approach to customizing the 3D scene representation, encouraging an interactive and highly configurable environment.

The Display Subpanel is dedicated to tracking and managing the 2D and 3D objects present within the scene. This Display Subpanel lists all the objects currently loaded into the system, enabling users to select, manipulate, and visualize each object independently. It provides visual cues, such as color coding for each 3D object shown in the 3D Scene Display, as an approach to differentiate between multiple objects and make it easier to identify the specific elements in the scene. Users can customize the transparency value of objects, which is helpful for visualizing cluttered and occluded areas or managing complex scenes with overlapping models.

By combining these three components, the Main Panel offers users complete control over the 6D pose annotation process. The Main Panel can also be hidden when displaying the 3D Scene as shown in Figure~\ref{fig:Vision6D_visualization_annotation} (a). Users can easily load data, apply camera modifications, and manipulate 3D objects within the GUI.

\subsubsection{3D Scene Display}
\begin{figure*}[htbp]
    \centering
    \begin{minipage}{0.24\textwidth}
        \centering
        \includegraphics[width=\linewidth]{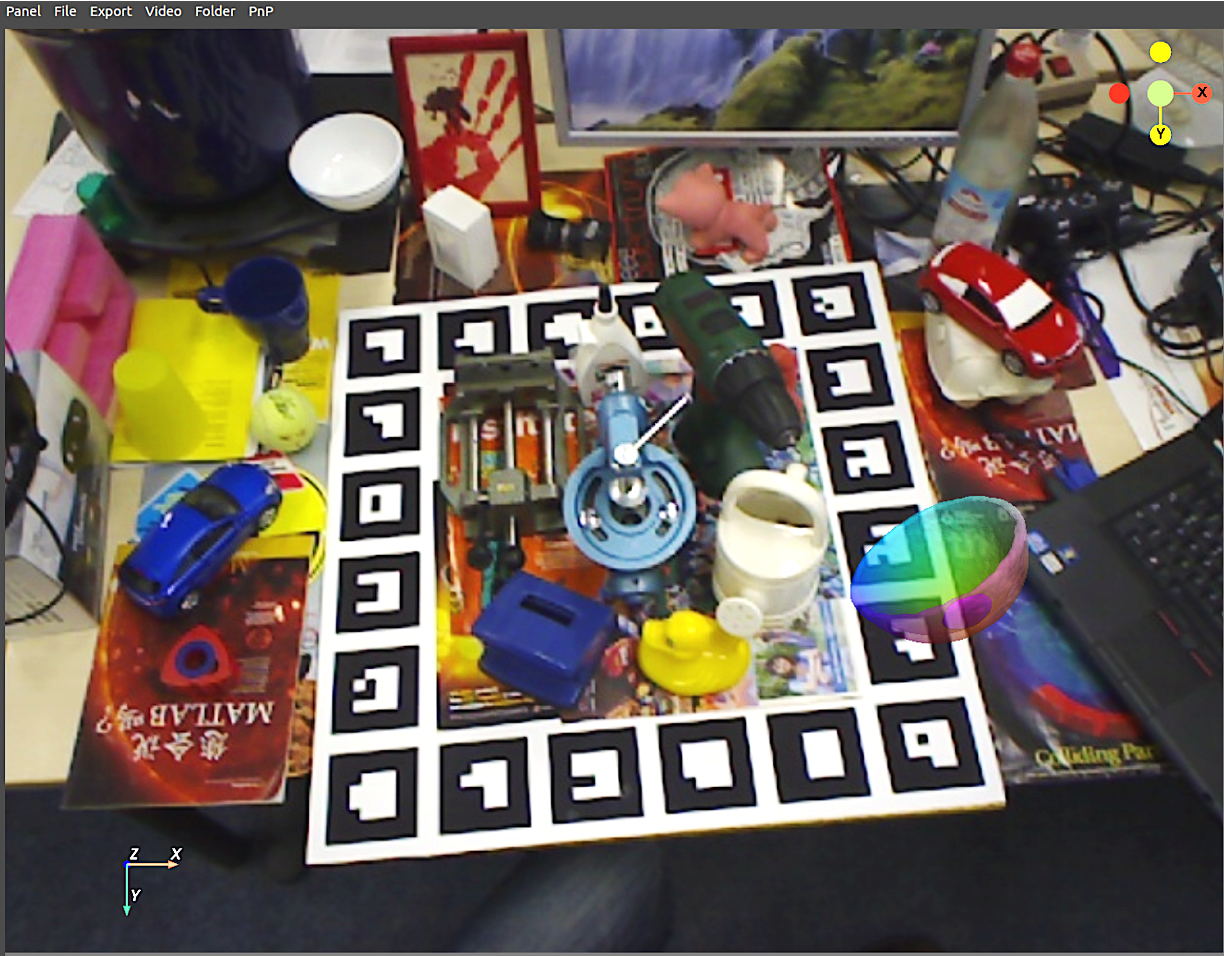}
        (a) Hide Main Panel
        \label{fig:hidden_main_panel}
    \end{minipage}
    \hfill
    \begin{minipage}{0.24\textwidth}
        \centering
        \includegraphics[width=\linewidth]{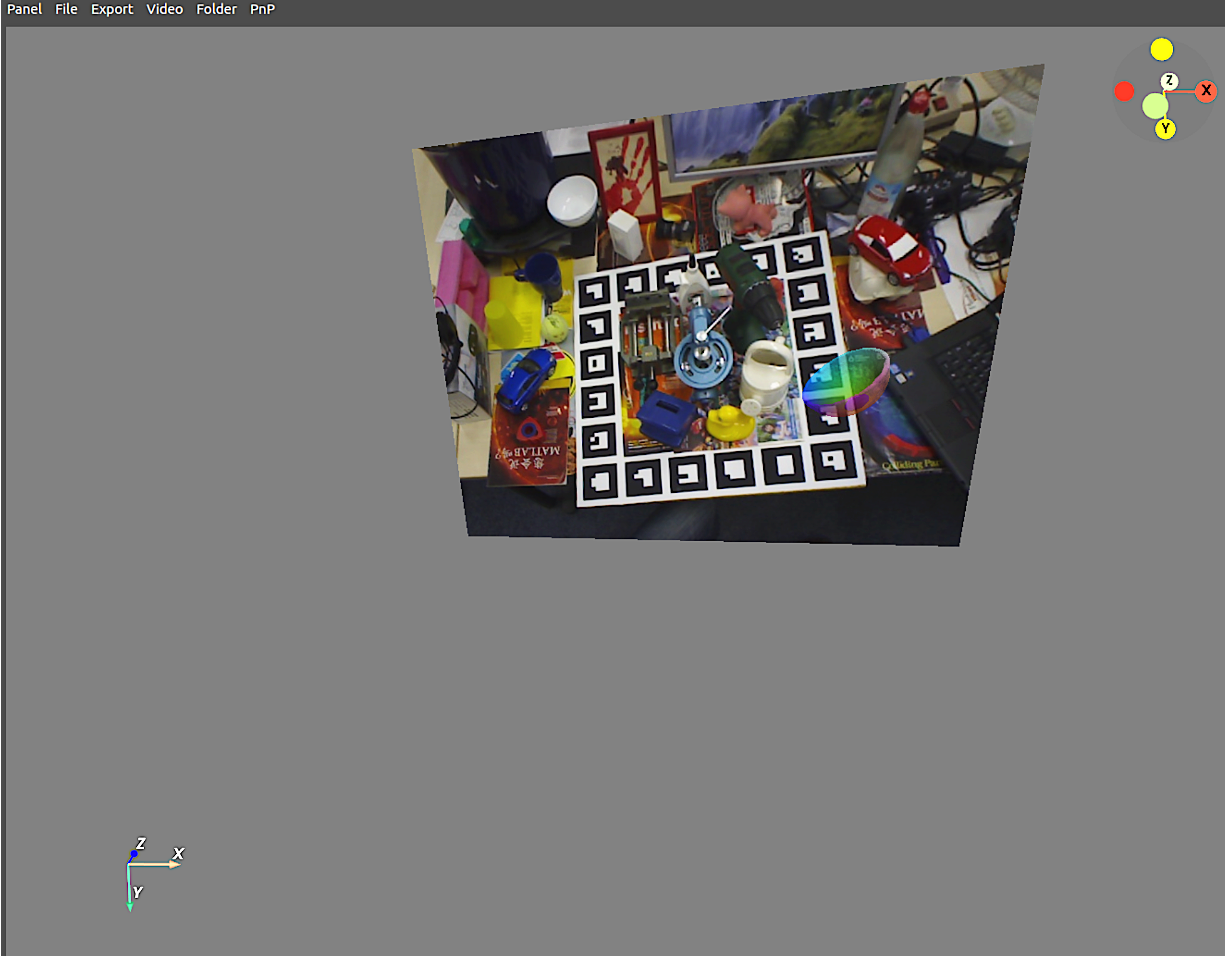}
        (b) Scene Camera
        \label{fig:scene_camera}
    \end{minipage}
    \hfill
    \begin{minipage}{0.24\textwidth}
        \centering
        \includegraphics[width=\linewidth]{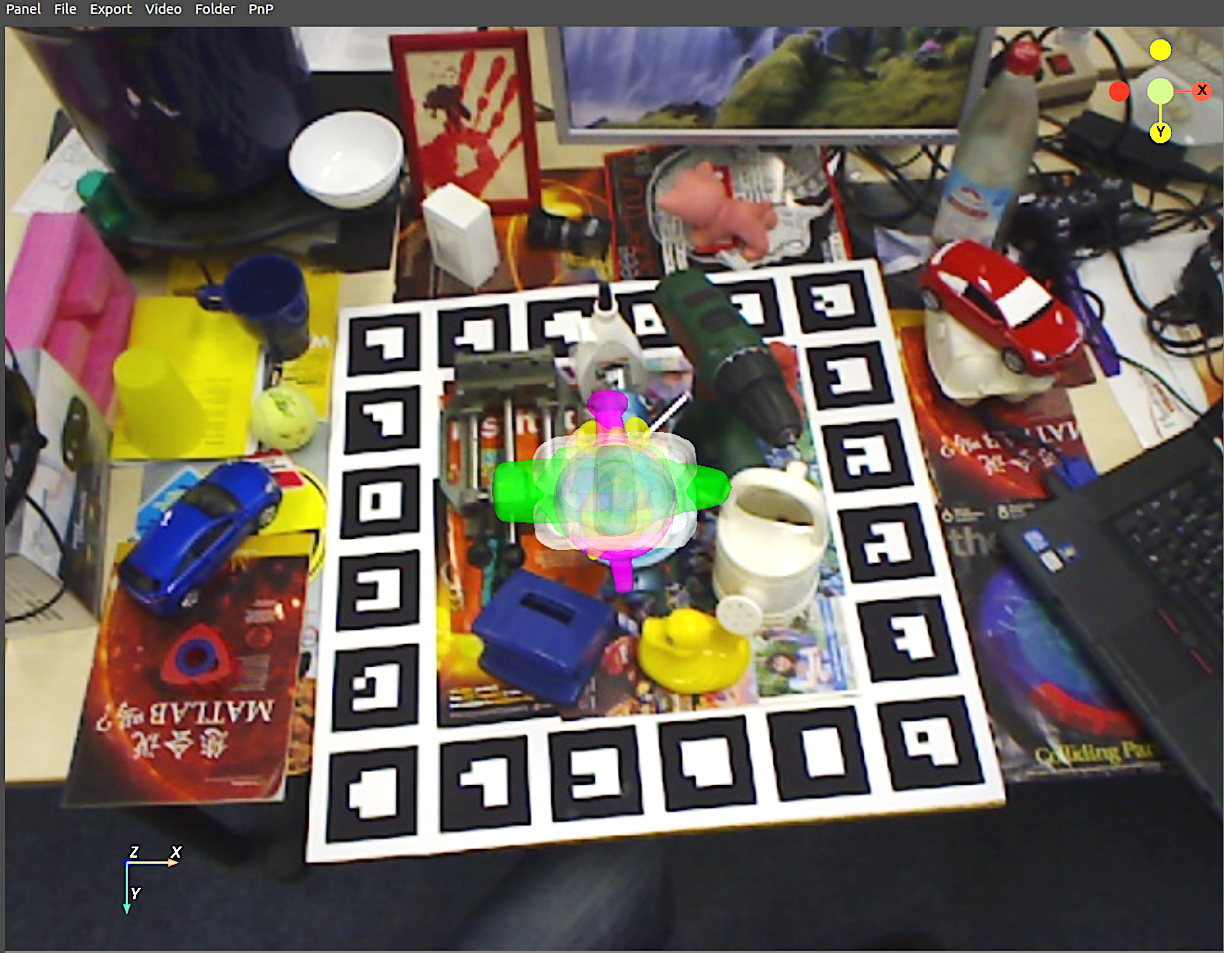}
        (c) Initial Pose
        \label{fig:initial_pose}
    \end{minipage}
    \hfill
    \begin{minipage}{0.24\textwidth}
        \centering
        \includegraphics[width=\linewidth]{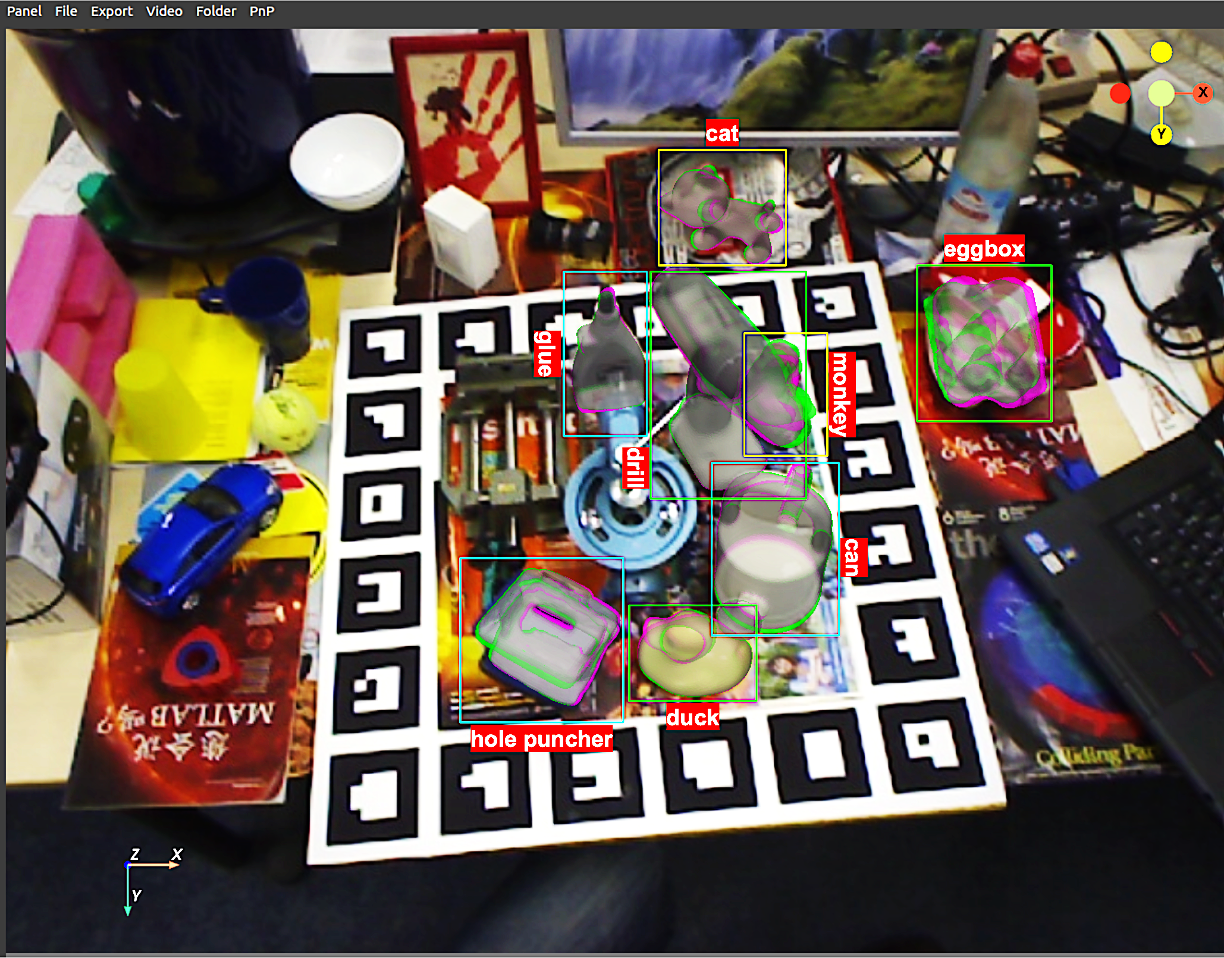}
        (d) Pose Comparison
        \label{fig:pose_comparison}
    \end{minipage}
    
    \caption{Visualization and Pose Annotation of Vision6D}
    \label{fig:Vision6D_visualization_annotation}
\end{figure*}
The 3D Scene Display serves as the primary workspace where users can interact with and display both 2D images and 3D models. It provides an interactive 3D scene that assists in the pose annotation process, showing the alignment of the 3D models and their respective 2D projections. Users can visually inspect the accuracy of the pose by observing how the 3D model is rendered onto the image scene. 
As the user adjusts the 3D object's pose within the scene with their mouse and keyboard, Vision6D updates the corresponding 2D projection immediately. This visual feedback allows users to quickly verify and refine the object's pose by comparing the rendered 3D model against the input 2D image. The accuracy of this correspondence is key to ensuring that the object's pose is correctly aligned with the camera's perspective, which is important for applications requiring high-precision pose annotations. 


Vision6D utilizes a multi-view system to provide a flexible user experience. The first viewpoint is the scene camera, which allows the user to freely navigate and explore the scene from different angles. This pose provides an independent view of the entire 3D environment as shown in Figure~\ref{fig:Vision6D_visualization_annotation} (b). In the top-right corner of Vision6D, users will find the scene camera orientation controls, a small visual representation of the camera axes $xyz$ in red, green, and yellow, respectively. These camera orientation controls allow users to quickly see the current orientation of the camera within the 3D space and adjust the camera view to standard perspectives, such as the top view, side view, or front view, with a single click. The camera orientation controls provide additional assistance when navigating the 3D scene, helping users maintain a sense of spatial awareness as they manipulate 3D objects. Switching or resetting these views allows for easier object alignment and more accurate judgment of annotated poses from various angles. This improves spatial understanding, especially when addressing the challenges of depth estimation.

Another camera view is the original camera that formulated the rendered scene image, which mimics the position and orientation of the camera that was used to capture the input image. This camera is essential for rendering the scene from the same perspective as the 2D image. The proposed multi-view system provides flexibility as users can interact with the scene and adjust the objects' poses while consistently comparing it to how it appears from the original camera's perspective.

A key feature in the 3D Scene Display is the ability to annotate object poses using intuitive click-and-drag gestures. When objects are first loaded into Vision6D, the objects' poses are set to an identity matrix as shown in Figure~\ref{fig:Vision6D_visualization_annotation} (c). Users can directly interact with 3D objects in the scene by selecting a target mesh via a mouse and adjusting their position or rotation by drag gestures. Moreover, users can also fine-tune the pose annotations by zooming in on specific parts of the scene for closer inspection. Using these control systems, Figure~\ref{fig:Vision6D_visualization_annotation} (d) shows the user-annotated poses for eight meshes colored in lime, and the ground-truth poses obtained from the Linemod-Occluded dataset colored in magenta. The rendered ground-truth and annotated poses closely match visually.

\subsubsection{Output Panel}
The Output Panel located at the bottom left of Vision6D is shown in Figure~\ref{fig:system_overview}. This panel is used to output the current annotated pose and display pose data history. It shows immediate feedback on any transformations applied to the 3D objects, demonstrating the current state of the objects' poses in a matrix form. Users can easily monitor changes, copy to the clipboard, and clear the Output Panel as they use Vision6D. 

\section{User Study}
\label{sec:user_study}
In this section, we present a user study that was conducted to compare and evaluate both the accuracy and the overall effectiveness of Vision6D in a manner that demonstrates its usability, intuitive design, and practical applicability for generating precise 6D pose annotations in various authentic scenarios.

\subsection{Hypotheses}
Specifically, our objective is to validate the following hypotheses using multiple evaluation metrics:
\begin{itemize}
    \item \textbf{H1: Annotation Accuracy Hypothesis.} Vision6D enables users to generate 6D pose annotations that are statistically comparable to the default ground-truth object poses from publicly available datasets. This hypothesis suggests that the tool produces results closely aligned with pre-existing ground-truth annotations captured directly by a camera. We use multiple metrics to measure positional and rotational offsets between Vision6D annotations and ground-truth poses through both inter-personal consistency and intra-personal variability evaluations.
    \item \textbf{H2: Annotation Efficiency Hypothesis.} The interactive 3D-to-2D interface of Vision6D aims to reduce the time and effort required for users to generate accurate 6D poses. We anticipate that the user interface's intuitive controls, direct manipulation features, and immediate feedback will ease the workflow, resulting in shorter annotation time without compromising accuracy. This is evaluated through annotation duration time analysis.
\end{itemize}
\begin{figure*}[!ht]
    \centering
    \begin{minipage}[t]{0.4\textwidth}
        \begin{minipage}[c]{0.05\textwidth}
            \centering
            \rotatebox{90}{Angular Distance (in degrees)}
        \end{minipage}
        \begin{minipage}{0.95\textwidth}
            \includegraphics[width=\textwidth]{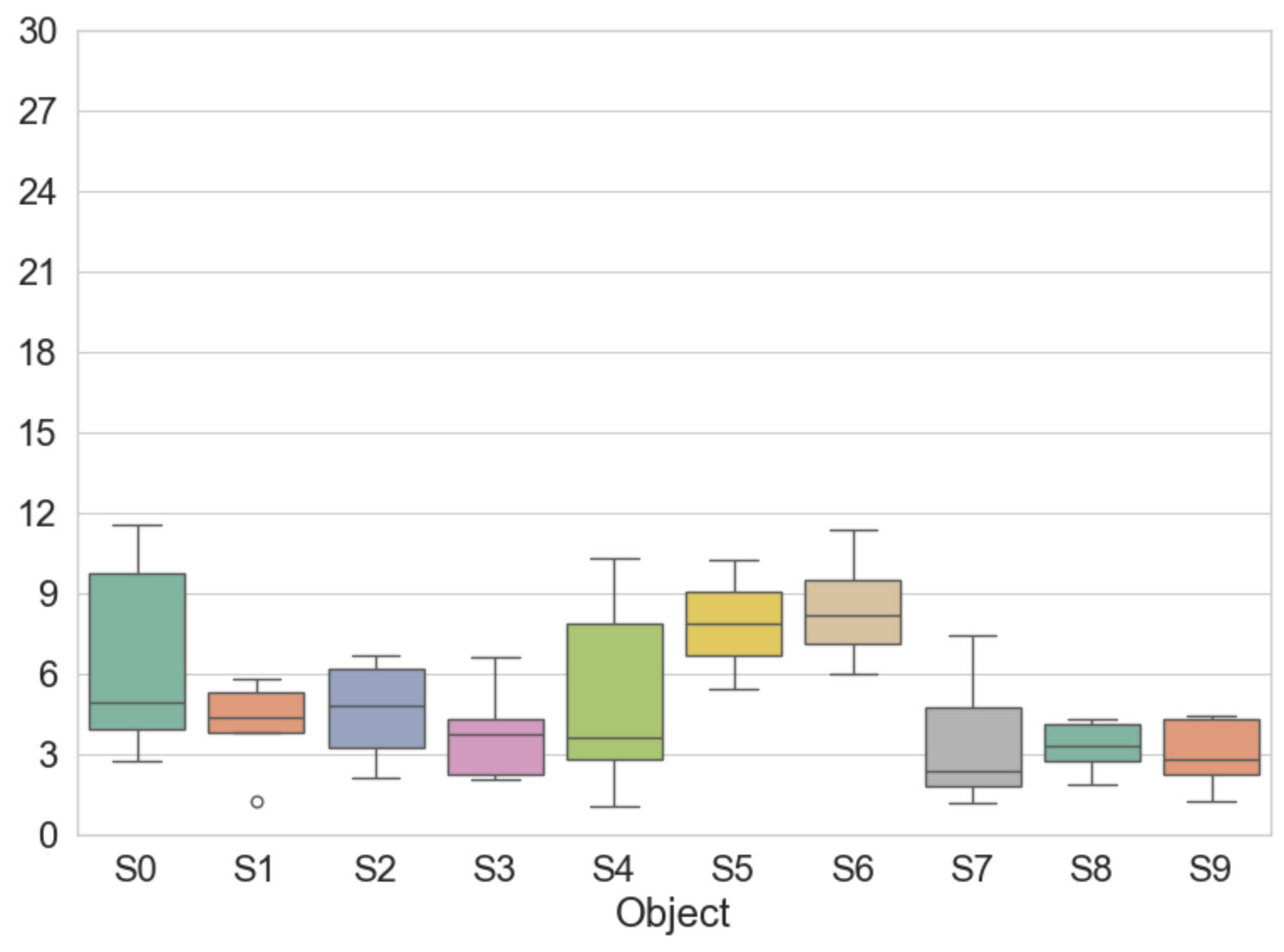}
        \end{minipage}\hspace{-0.3em}
    \end{minipage}
    \hspace{3em}
    \begin{minipage}[t]{0.4\textwidth}
        \begin{minipage}[c]{0.05\textwidth}
            \centering
            \rotatebox{90}{Angular Distance (in degrees)}
        \end{minipage}
        \begin{minipage}{0.95\textwidth}
            \includegraphics[width=\textwidth]{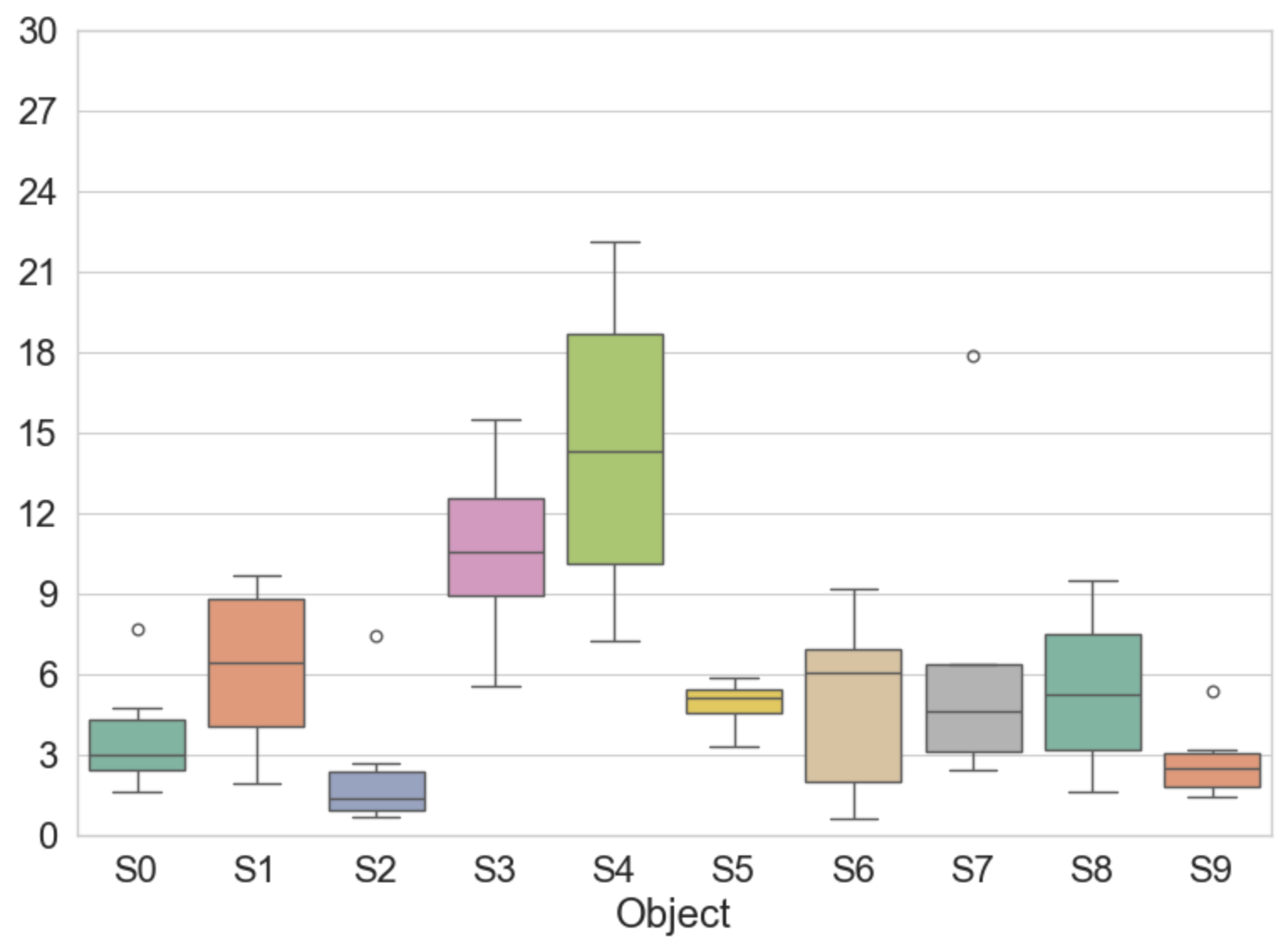}
        \end{minipage}\hspace{-0.3em}
    \end{minipage}

    \begin{minipage}[t]{0.4\textwidth}
        \begin{minipage}[c]{0.05\textwidth}
            \centering
            \rotatebox{90}{Euclidean Distance (in mm)}
        \end{minipage}
        \begin{minipage}{0.95\textwidth}
            \includegraphics[width=\textwidth]{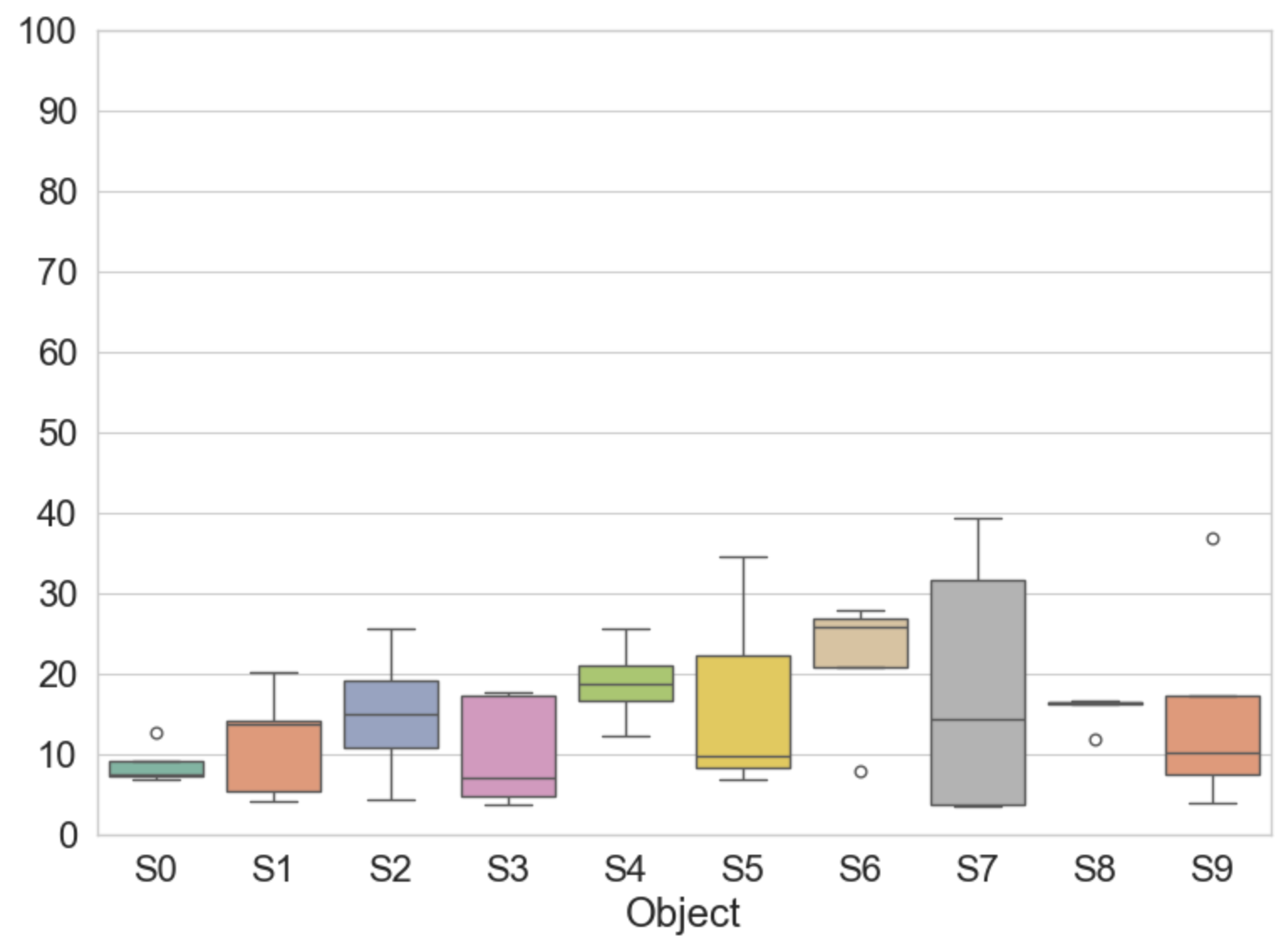}
        \end{minipage}\hspace{-0.3em}
    \end{minipage}
    \hspace{3em}
    \begin{minipage}[t]{0.4\textwidth}
        \begin{minipage}[c]{0.05\textwidth}
            \centering
            \rotatebox{90}{Euclidean Distance (in mm)}
        \end{minipage}
        \begin{minipage}{0.95\textwidth}
            \includegraphics[width=\textwidth]{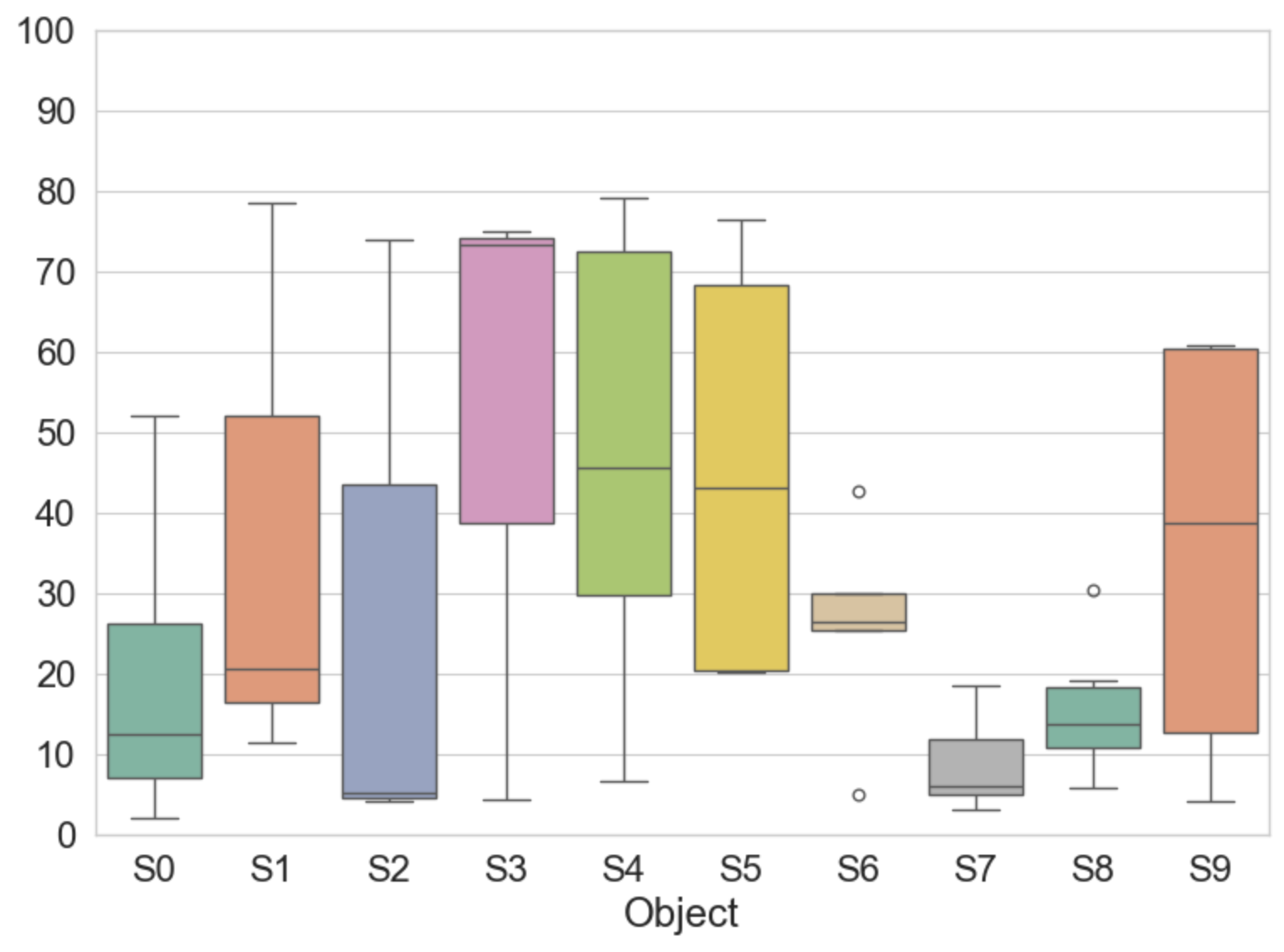}
        \end{minipage}\hspace{-0.3em}
    \end{minipage}

    \begin{minipage}[t]{0.4\textwidth}
        \begin{minipage}[c]{0.05\textwidth}
            \centering
            \rotatebox{90}{ADD Metric (in mm)}
        \end{minipage}
        \begin{minipage}{0.95\textwidth}
            \includegraphics[width=\textwidth]{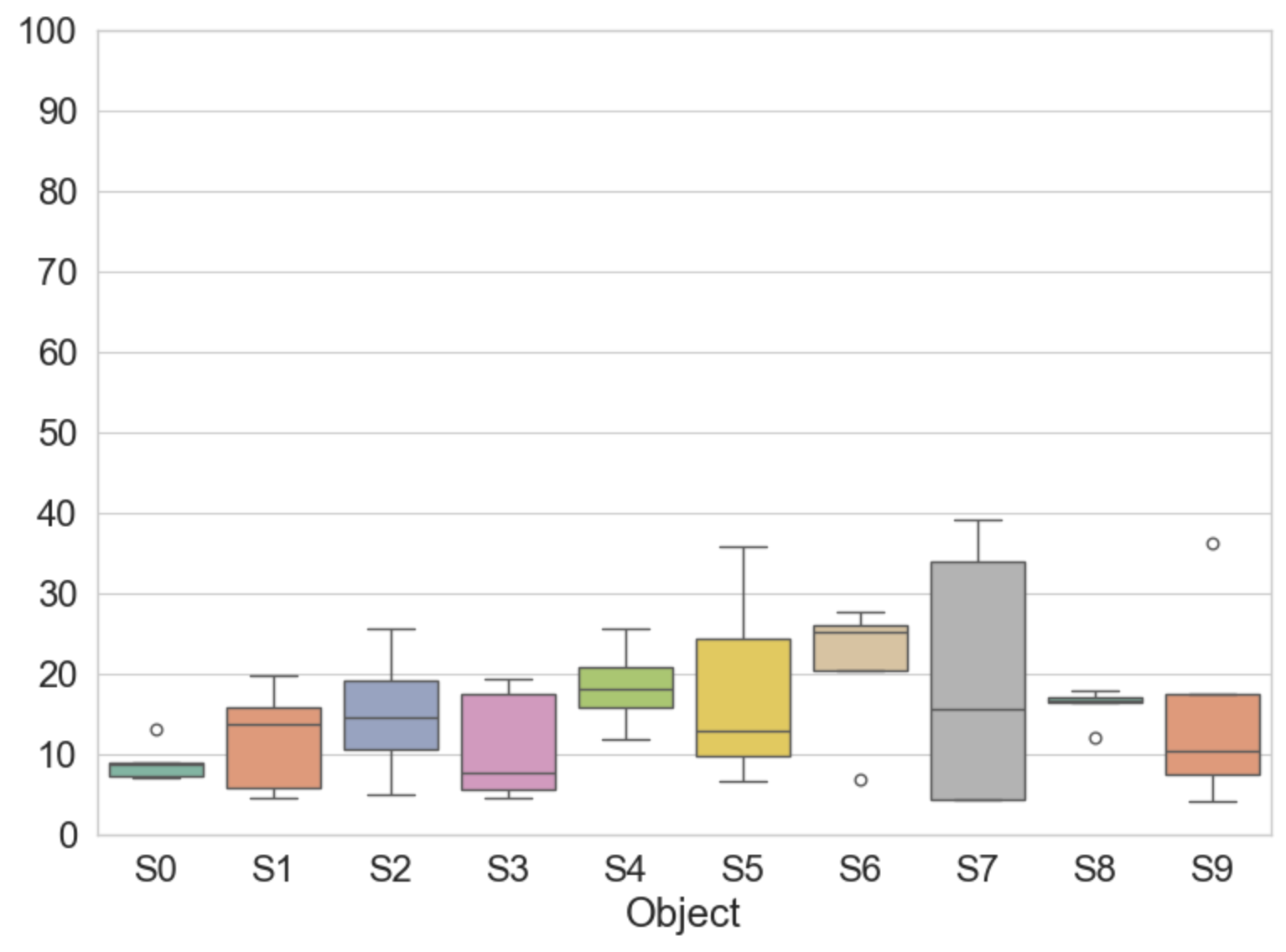}
        \end{minipage}\hspace{-0.3em}

        \centering
        Linemod Dataset
    \end{minipage}
    \hspace{3em}
    \begin{minipage}[t]{0.4\textwidth}
        \begin{minipage}[c]{0.05\textwidth}
            \centering
            \rotatebox{90}{ADD Metric (in mm)}
        \end{minipage}
        \begin{minipage}{0.95\textwidth}
            \includegraphics[width=\textwidth]{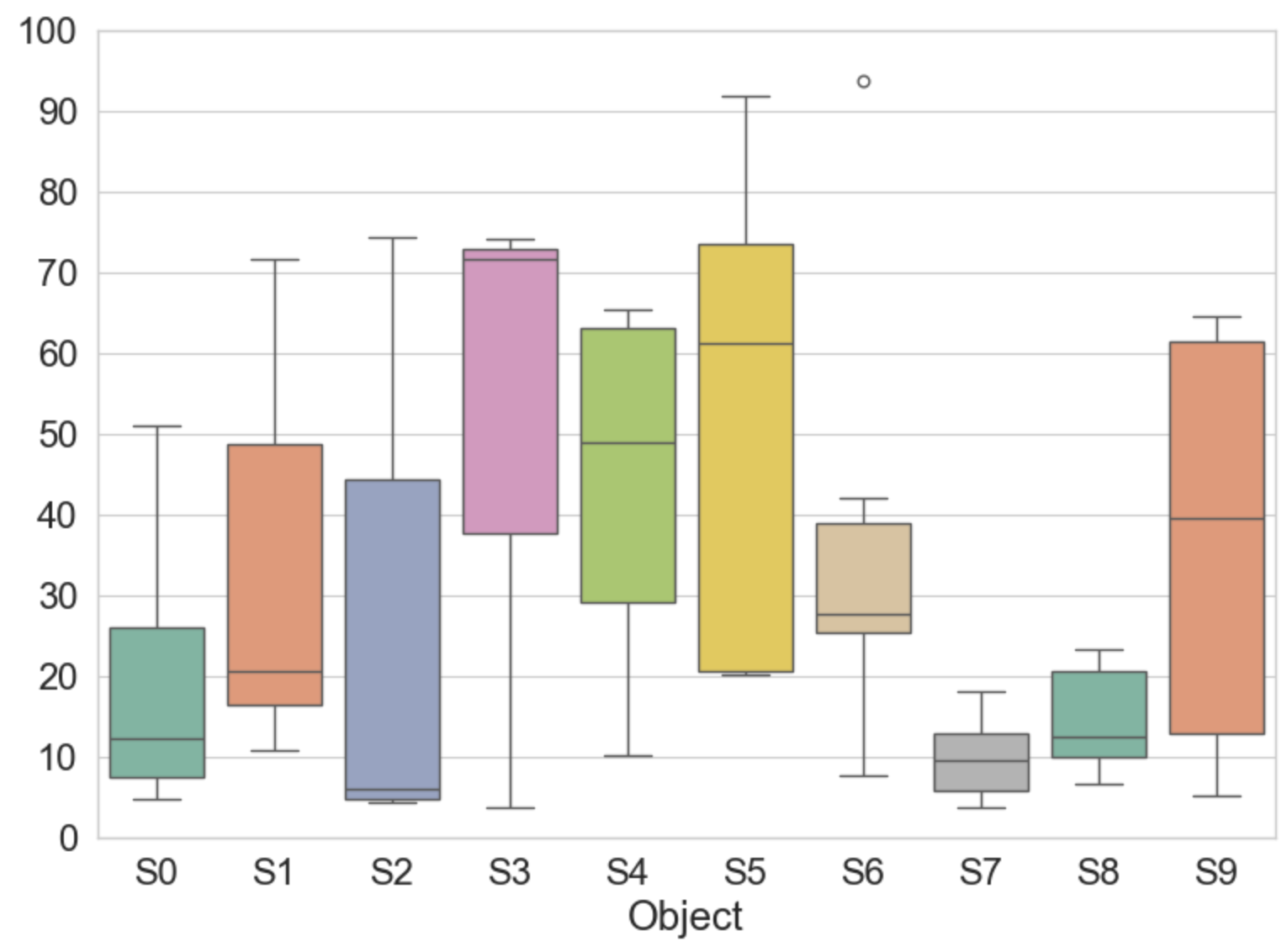}
        \end{minipage}\hspace{-0.3em}
        
        \centering
        HANDAL Dataset
    \end{minipage}
    
    \caption{\textbf{Quantitative Evaluation of Inter-Personal Variability.} This figure presents the distributions of Angular Distance, Euclidean Distance, and ADD metrics for assessing inter-personal variability. The results demonstrate the effectiveness of pose annotation using Vision6D.}
    \label{fig:inter_personal_consistency}
\end{figure*}

\begin{figure*}[!ht]
     \centering
    \begin{minipage}[t]{0.4\textwidth}
        \begin{minipage}[c]{0.05\textwidth}
            \centering
            \rotatebox{90}{Angular Distance (in degrees)}
        \end{minipage}
        \begin{minipage}{0.95\textwidth}
            \includegraphics[width=\textwidth]{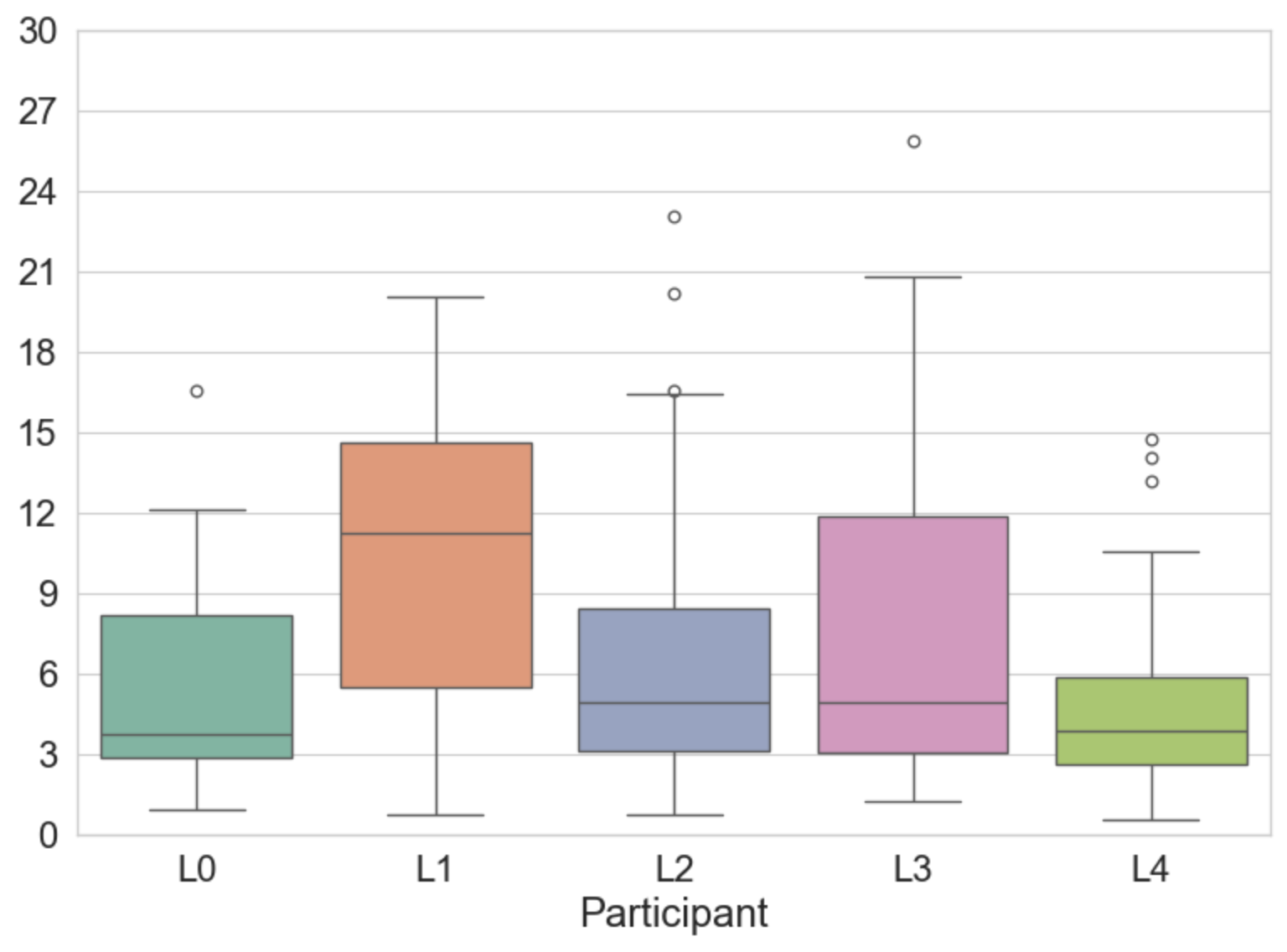}
        \end{minipage}\hspace{-0.3em}
    \end{minipage}
    \hspace{3em}
    \begin{minipage}[t]{0.4\textwidth}
        \begin{minipage}[c]{0.05\textwidth}
            \centering
            \rotatebox{90}{Angular Distance (in degrees)}
        \end{minipage}
        \begin{minipage}{0.95\textwidth}
            \includegraphics[width=\textwidth]{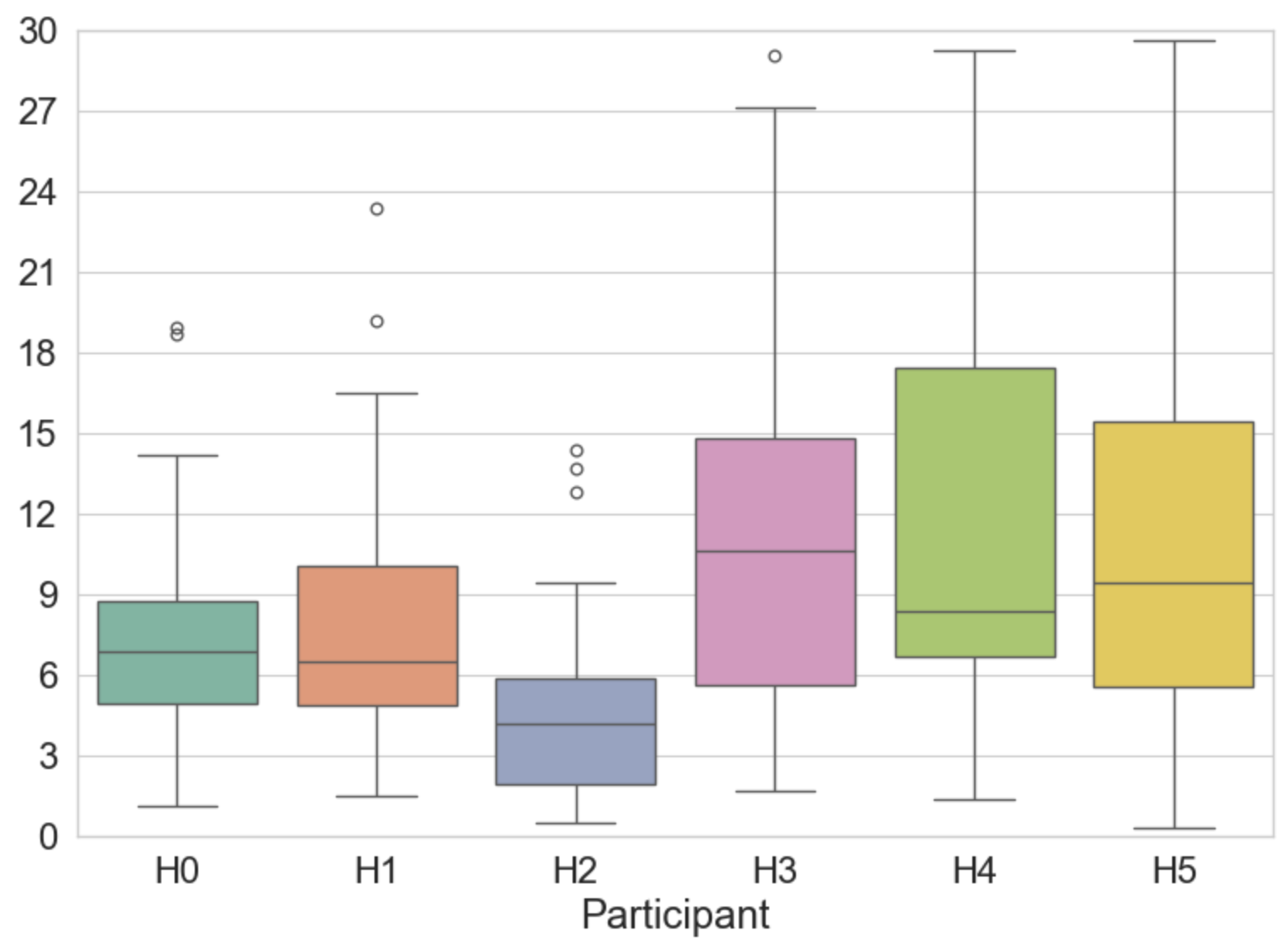}
        \end{minipage}\hspace{-0.3em}
    \end{minipage}

    \begin{minipage}[t]{0.4\textwidth}
        \begin{minipage}[c]{0.05\textwidth}
            \centering
            \rotatebox{90}{Euclidean Distance (in mm)}
        \end{minipage}
        \begin{minipage}{0.95\textwidth}
            \includegraphics[width=\textwidth]{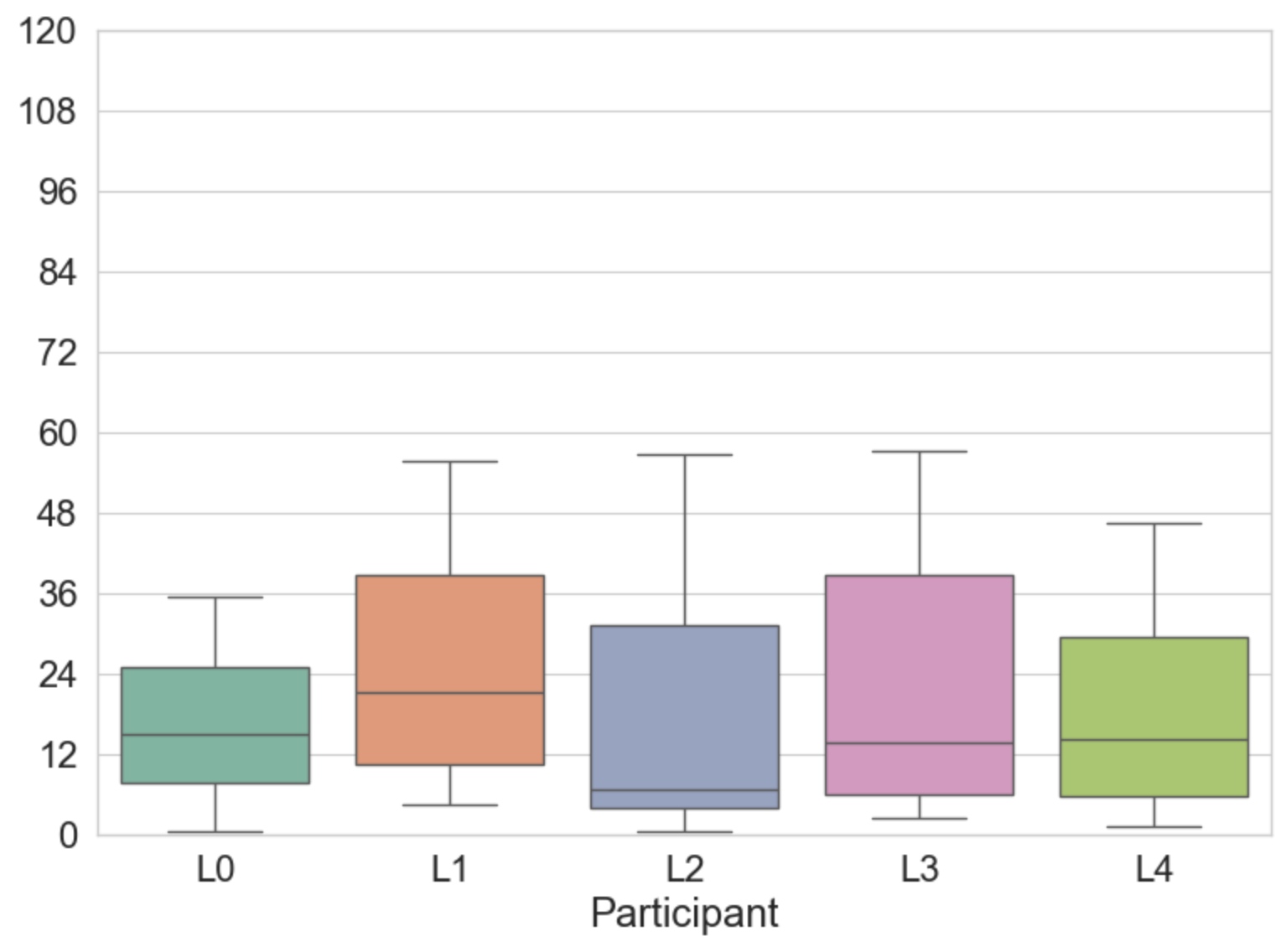}
        \end{minipage}\hspace{-0.3em}
    \end{minipage}
    \hspace{3em}
    \begin{minipage}[t]{0.4\textwidth}
        \begin{minipage}[c]{0.05\textwidth}
            \centering
            \rotatebox{90}{Euclidean Distance (in mm)}
        \end{minipage}
        \begin{minipage}{0.95\textwidth}
            \includegraphics[width=\textwidth]{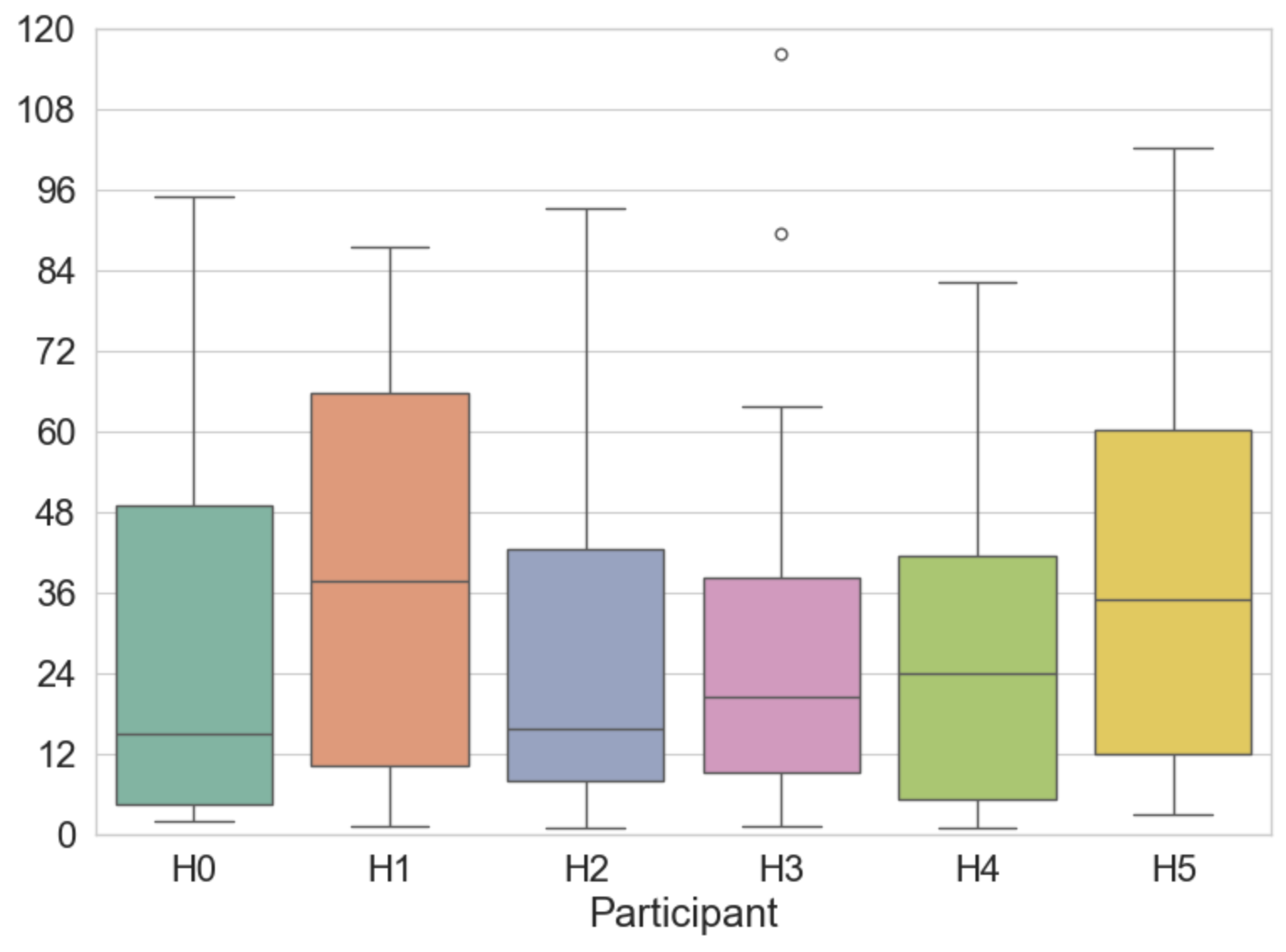}
        \end{minipage}\hspace{-0.3em}
    \end{minipage}

    \begin{minipage}[t]{0.4\textwidth}
        \begin{minipage}[c]{0.05\textwidth}
            \centering
            \rotatebox{90}{ADD Metric (in mm)}
        \end{minipage}
        \begin{minipage}{0.95\textwidth}
            \includegraphics[width=\textwidth]{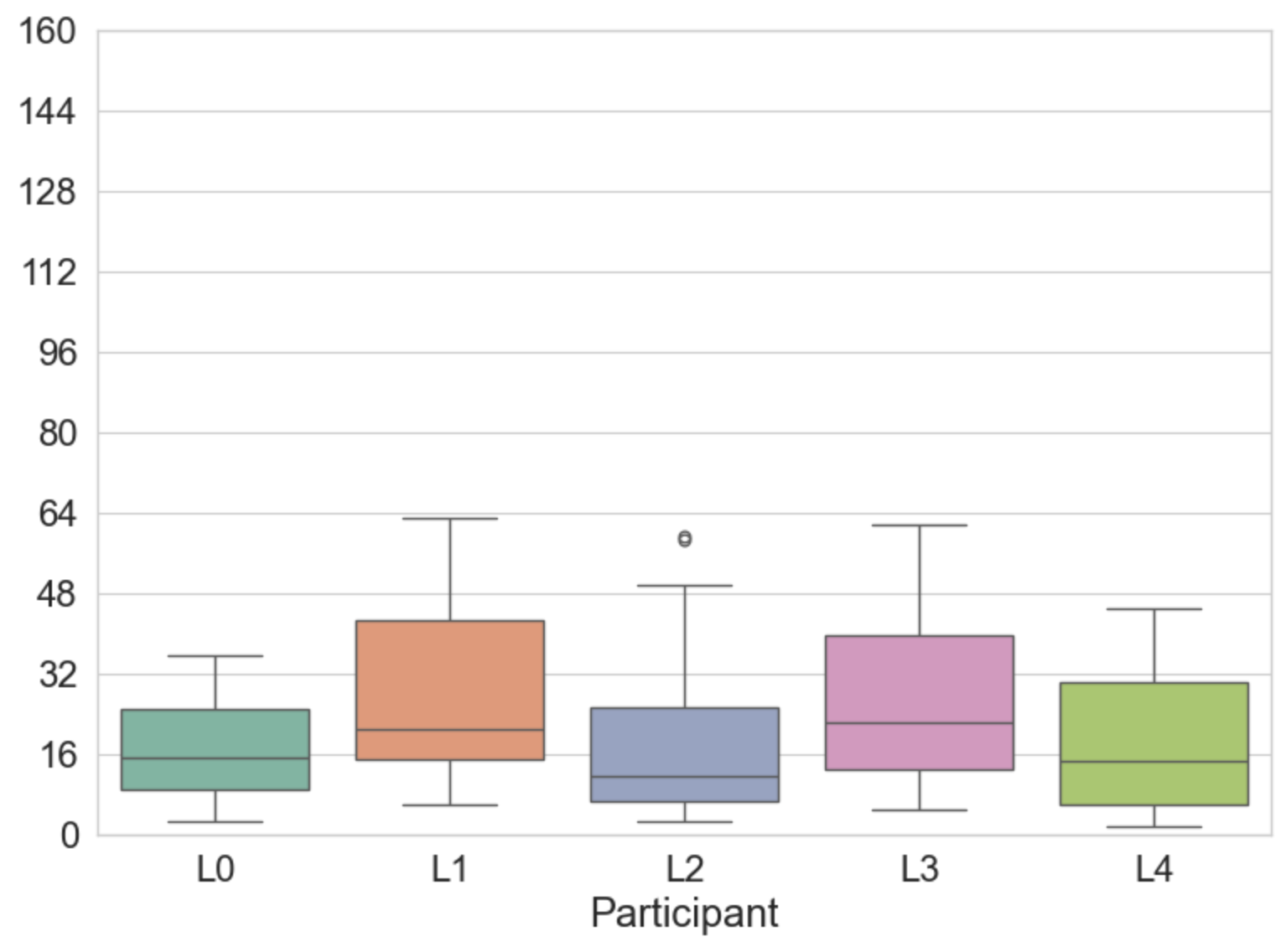}
        \end{minipage}\hspace{-0.3em}

        \centering
        Linemod Dataset
    \end{minipage}
    \hspace{3em}
    \begin{minipage}[t]{0.4\textwidth}
        \begin{minipage}[c]{0.05\textwidth}
            \centering
            \rotatebox{90}{ADD Metric (in mm)}
        \end{minipage}
        \begin{minipage}{0.95\textwidth}
            \includegraphics[width=\textwidth]{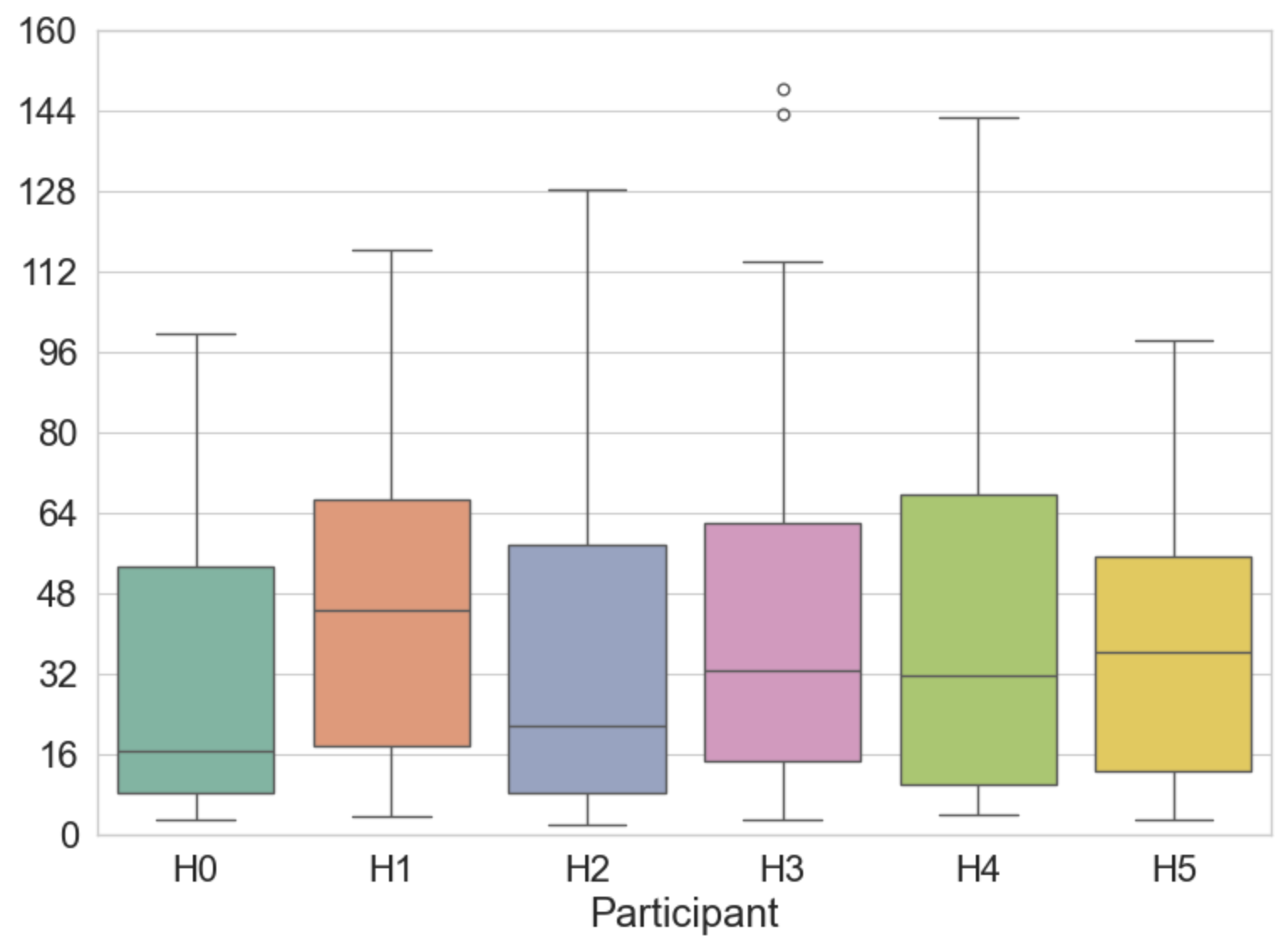}
        \end{minipage}\hspace{-0.3em}
        
        \centering
        HANDAL Dataset
    \end{minipage}
    
    \caption{\textbf{Quantitative Evaluation of Intra-Personal Consistency.} This figure presents the distributions of Angular Distance, Euclidean Distance, and ADD metrics for assessing intra-personal consistency. The results highlight the robustness and reproducibility of the Vision6D.}
    \label{fig:intra_personal_consistency}
\end{figure*}

\begin{figure*}[ht]
    \begin{minipage}[t]{0.45\textwidth}
        \begin{minipage}[c]{0.05\textwidth}
            \centering
            \rotatebox{90}{Monkey}
        \end{minipage}
        \begin{minipage}{0.22\textwidth}
            \includegraphics[width=\textwidth]{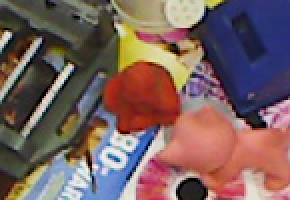}
        \end{minipage}
        \begin{minipage}{0.22\textwidth}
            \includegraphics[width=\textwidth]{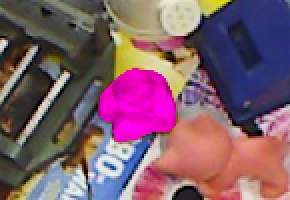}
        \end{minipage}
        \begin{minipage}{0.22\textwidth}
            \includegraphics[width=\textwidth]{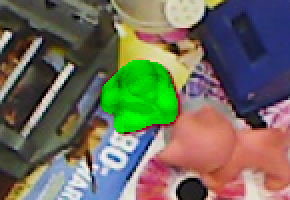}
        \end{minipage}
        \begin{minipage}{0.22\textwidth}
            \includegraphics[width=\textwidth]{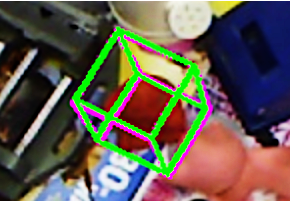}
        \end{minipage}
    \end{minipage}
    \hspace{2em}
    \vspace{1em}
    \begin{minipage}[t]{0.45\textwidth}
        \begin{minipage}[c]{0.05\textwidth}
            \centering
            \rotatebox{90}{Hammer1}
        \end{minipage}
        \begin{minipage}{0.22\textwidth}
            \includegraphics[width=\textwidth]{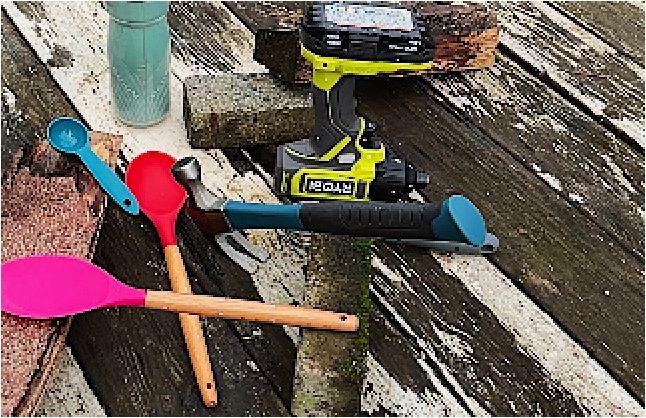}
        \end{minipage}
        \begin{minipage}{0.22\textwidth}
            \includegraphics[width=\textwidth]{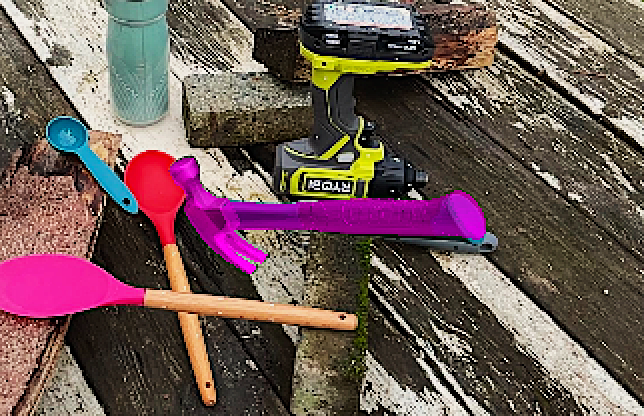}
        \end{minipage}
        \begin{minipage}{0.22\textwidth}
            \includegraphics[width=\textwidth]{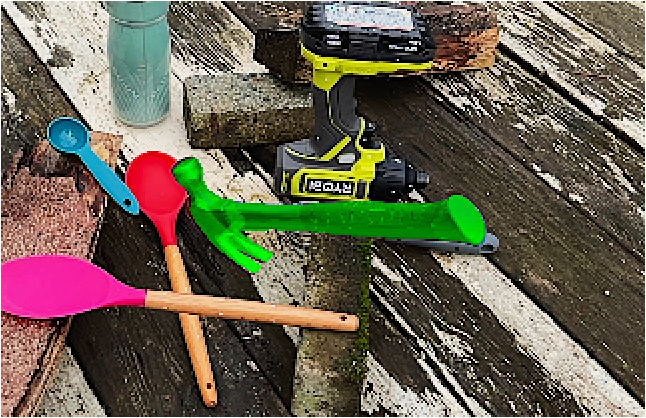}
        \end{minipage}
        \begin{minipage}{0.22\textwidth}
            \includegraphics[width=\textwidth]{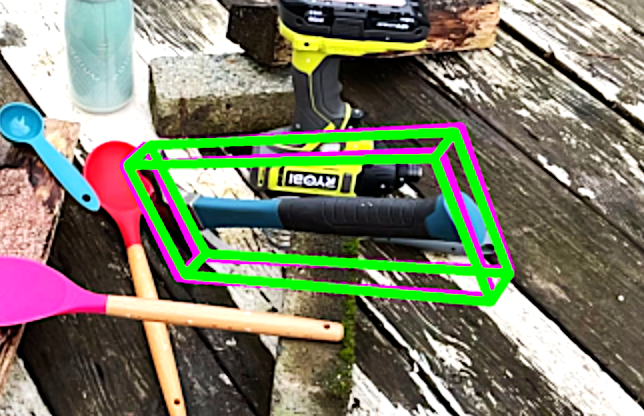}
        \end{minipage}
    \end{minipage}
    
    \begin{minipage}[t]{0.45\textwidth}
        \begin{minipage}[c]{0.05\textwidth}
            \centering
            \rotatebox{90}{Benchvice}
        \end{minipage}
        \begin{minipage}{0.22\textwidth}
            \includegraphics[width=\textwidth]{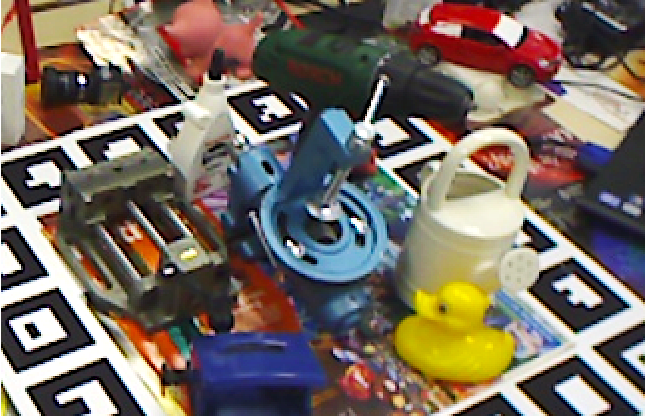}
        \end{minipage}
        \begin{minipage}{0.22\textwidth}
            \includegraphics[width=\textwidth]{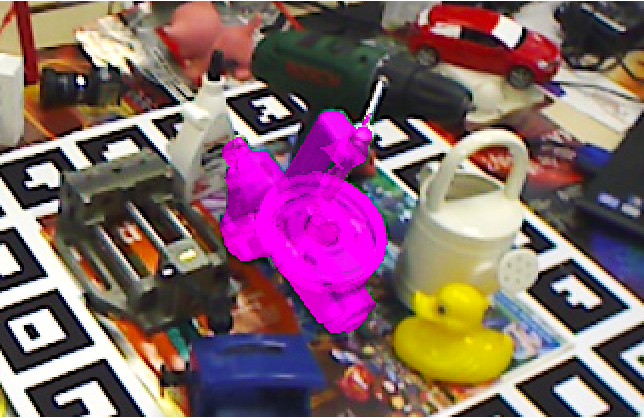}
        \end{minipage}
        \begin{minipage}{0.22\textwidth}
            \includegraphics[width=\textwidth]{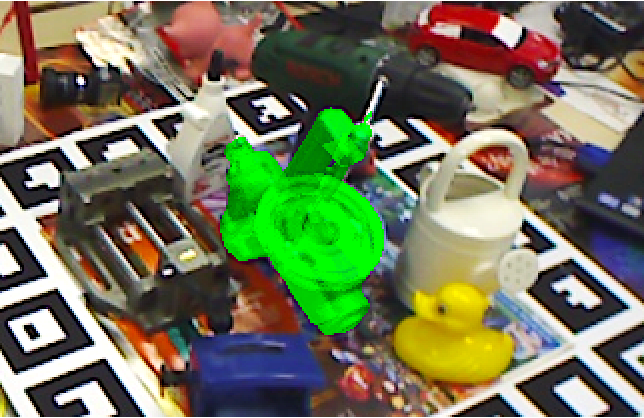}
        \end{minipage}
        \begin{minipage}{0.22\textwidth}
            \includegraphics[width=\textwidth]{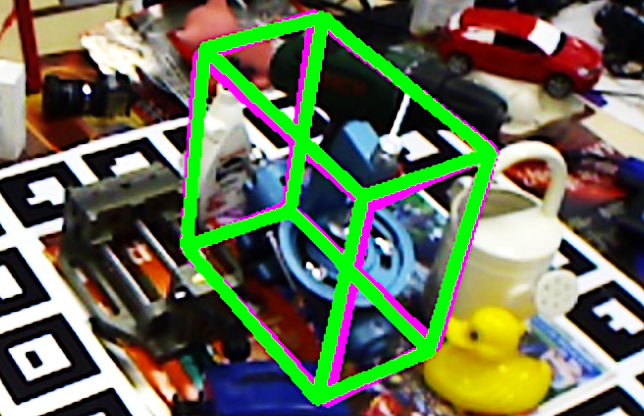}
        \end{minipage}
    \end{minipage}
    \hspace{2em}
    \vspace{1em}
    \begin{minipage}[t]{0.45\textwidth}
        \begin{minipage}[c]{0.05\textwidth}
            \centering
            \rotatebox{90}{Hammer2}
        \end{minipage}
        \begin{minipage}{0.22\textwidth}
            \includegraphics[width=\textwidth]{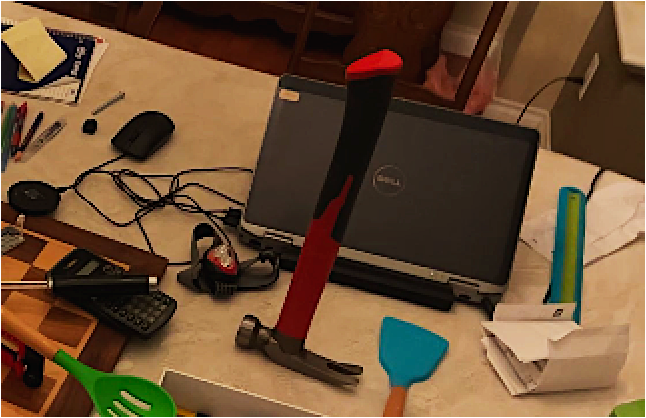}
        \end{minipage}
        \begin{minipage}{0.22\textwidth}
            \includegraphics[width=\textwidth]{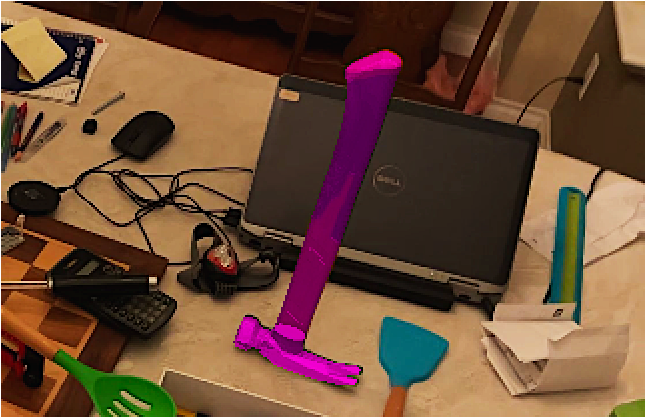}
        \end{minipage}
        \begin{minipage}{0.22\textwidth}
            \includegraphics[width=\textwidth]{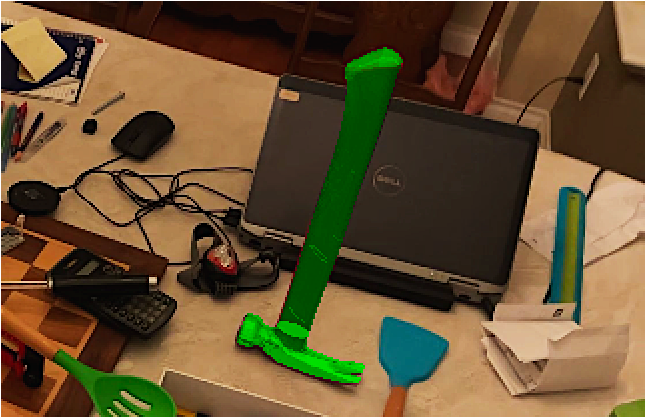}
        \end{minipage}
        \begin{minipage}{0.22\textwidth}
            \includegraphics[width=\textwidth]{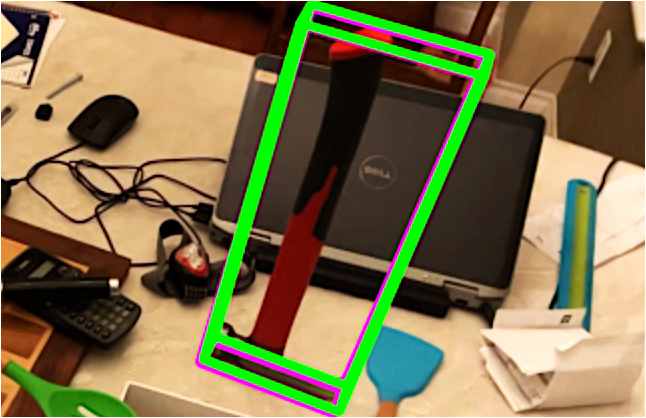}
        \end{minipage}
    \end{minipage}

    \begin{minipage}[t]{0.45\textwidth}
        \begin{minipage}[c]{0.05\textwidth}
            \centering
            \rotatebox{90}{Camera}
        \end{minipage}
        \begin{minipage}{0.22\textwidth}
            \includegraphics[width=\textwidth]{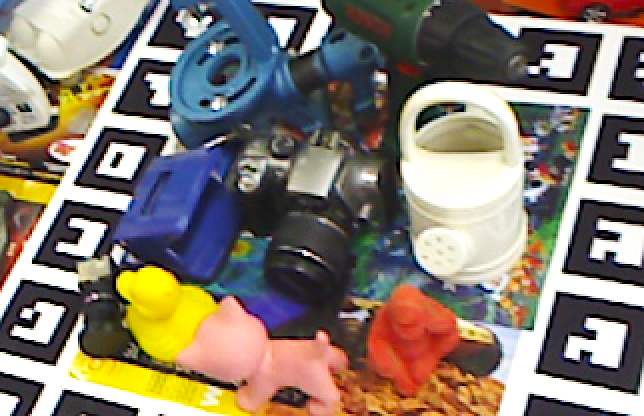}
        \end{minipage}
        \begin{minipage}{0.22\textwidth}
            \includegraphics[width=\textwidth]{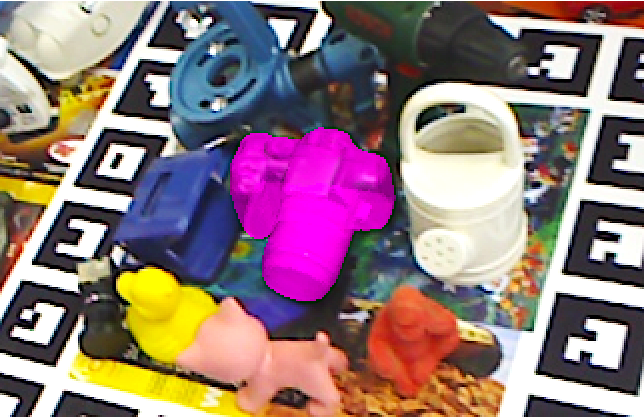}
        \end{minipage}
        \begin{minipage}{0.22\textwidth}
            \includegraphics[width=\textwidth]{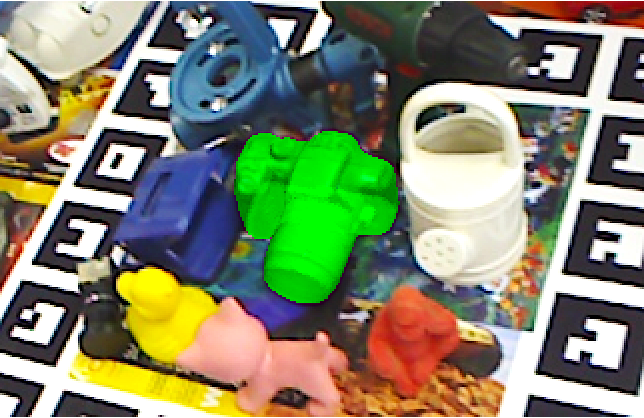}
        \end{minipage}
        \begin{minipage}{0.22\textwidth}
            \includegraphics[width=\textwidth]{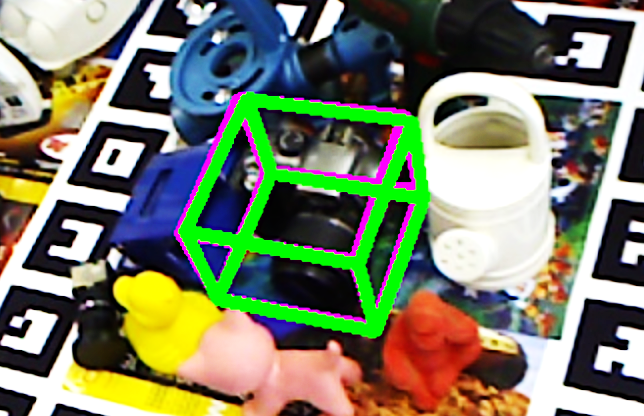}
        \end{minipage}
    \end{minipage}
    \hspace{2em}
    \vspace{1em}
    \begin{minipage}[t]{0.45\textwidth}
        \begin{minipage}[c]{0.05\textwidth}
            \centering
            \rotatebox{90}{Hammer3}
        \end{minipage}
        \begin{minipage}{0.22\textwidth}
            \includegraphics[width=\textwidth]{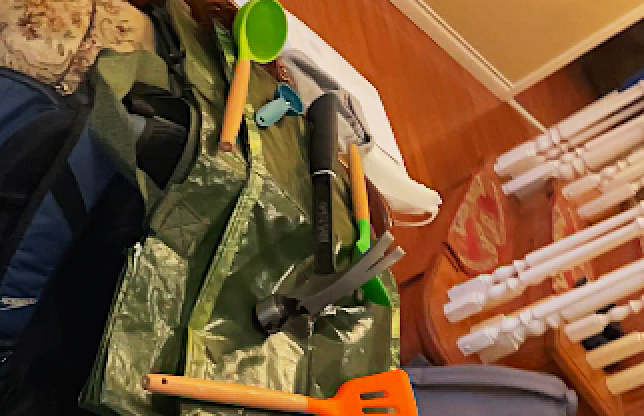}
        \end{minipage}
        \begin{minipage}{0.22\textwidth}
            \includegraphics[width=\textwidth]{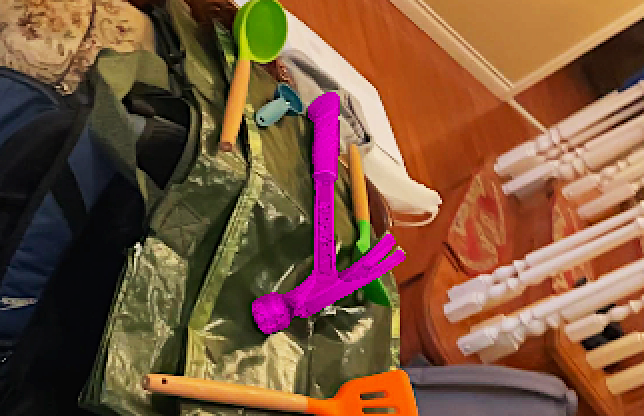}
        \end{minipage}
        \begin{minipage}{0.22\textwidth}
            \includegraphics[width=\textwidth]{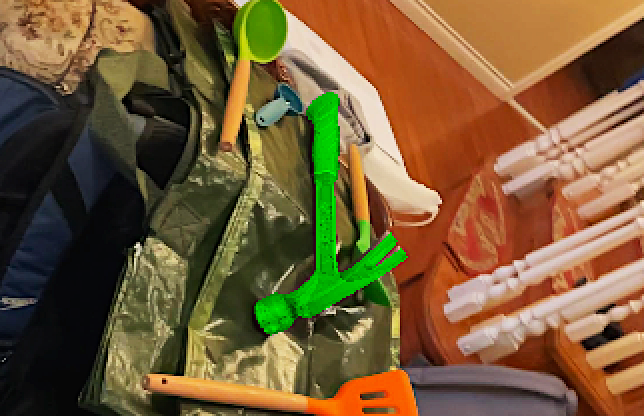}
        \end{minipage}
        \begin{minipage}{0.22\textwidth}
            \includegraphics[width=\textwidth]{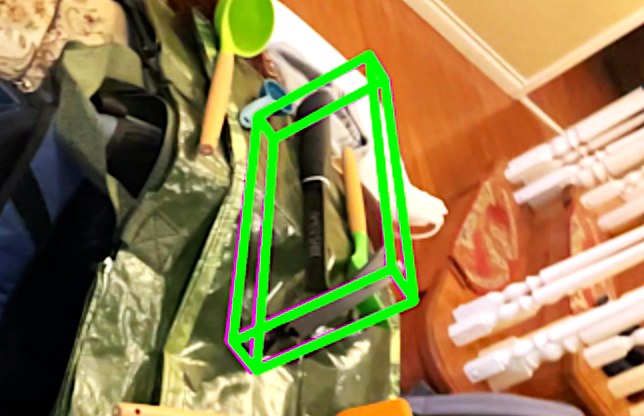}
        \end{minipage}
    \end{minipage}

    \begin{minipage}[t]{0.45\textwidth}
        \begin{minipage}[c]{0.05\textwidth}
            \centering
            \rotatebox{90}{Can}
        \end{minipage}
        \begin{minipage}{0.22\textwidth}
            \includegraphics[width=\textwidth]{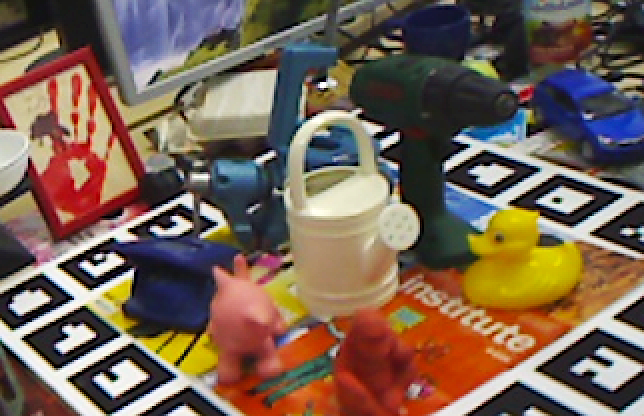}
        \end{minipage}
        \begin{minipage}{0.22\textwidth}
            \includegraphics[width=\textwidth]{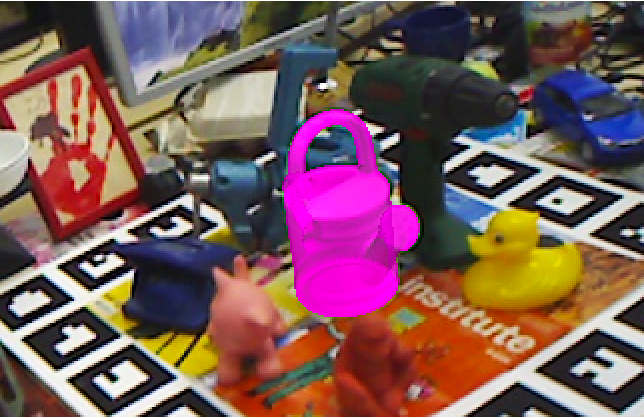}
        \end{minipage}
        \begin{minipage}{0.22\textwidth}
            \includegraphics[width=\textwidth]{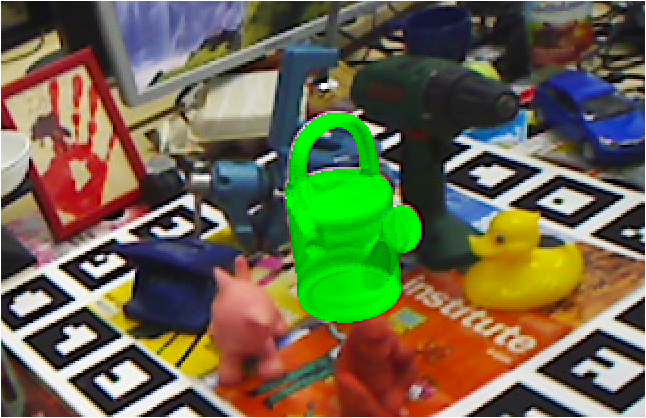}
        \end{minipage}
        \begin{minipage}{0.22\textwidth}
            \includegraphics[width=\textwidth]{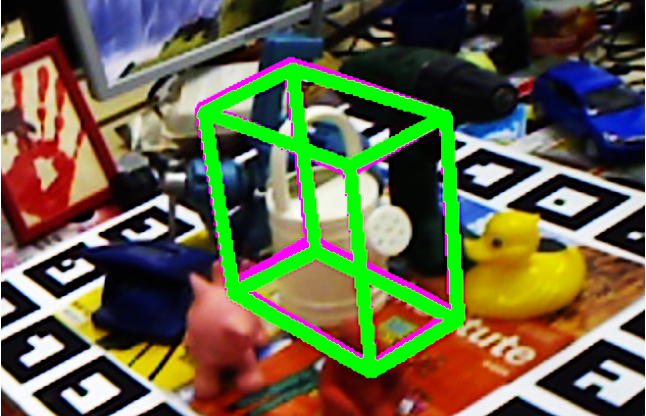}
        \end{minipage}
    \end{minipage}
    \hspace{2em}
    \vspace{1em}
    \begin{minipage}[t]{0.45\textwidth}
        \begin{minipage}[c]{0.05\textwidth}
            \centering
            \rotatebox{90}{Utensil1}
        \end{minipage}
        \begin{minipage}{0.22\textwidth}
            \includegraphics[width=\textwidth]{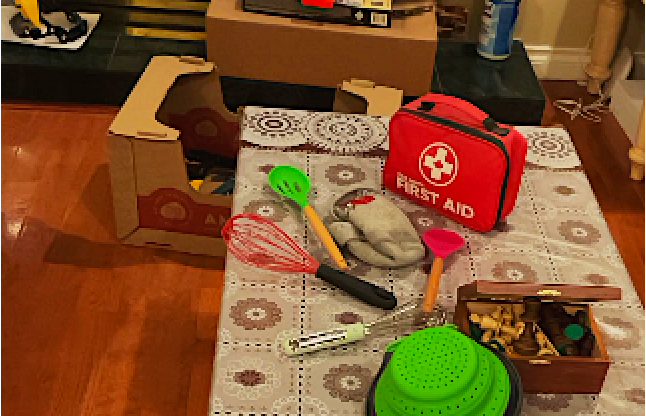}
        \end{minipage}
        \begin{minipage}{0.22\textwidth}
            \includegraphics[width=\textwidth]{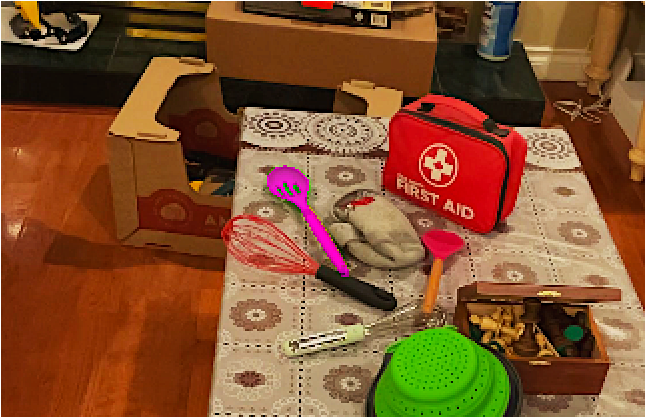}
        \end{minipage}
        \begin{minipage}{0.22\textwidth}
            \includegraphics[width=\textwidth]{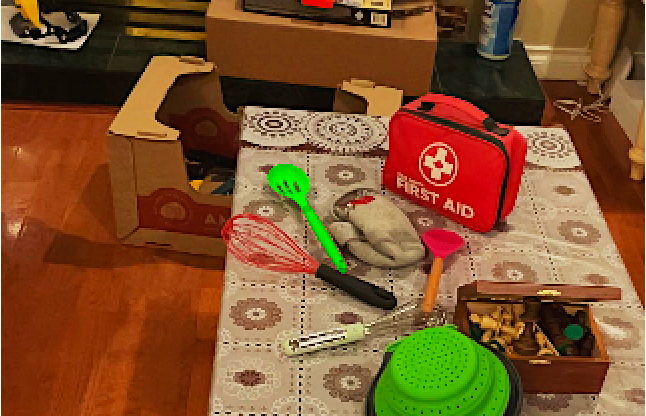}
        \end{minipage}
        \begin{minipage}{0.22\textwidth}
            \includegraphics[width=\textwidth]{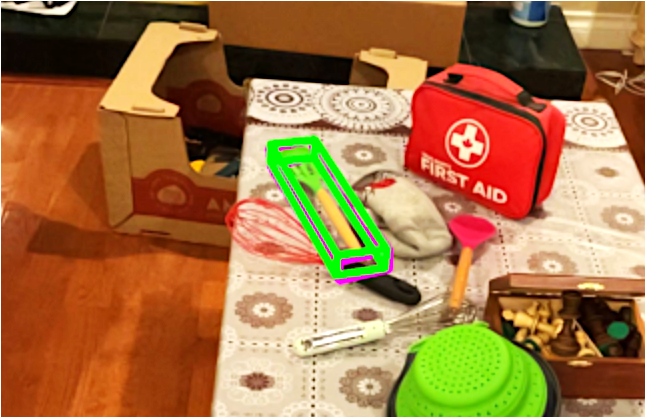}
        \end{minipage}
    \end{minipage}

    \begin{minipage}[t]{0.45\textwidth}
        \begin{minipage}[c]{0.05\textwidth}
            \centering
            \rotatebox{90}{Mug}
        \end{minipage}
        \begin{minipage}{0.22\textwidth}
            \includegraphics[width=\textwidth]{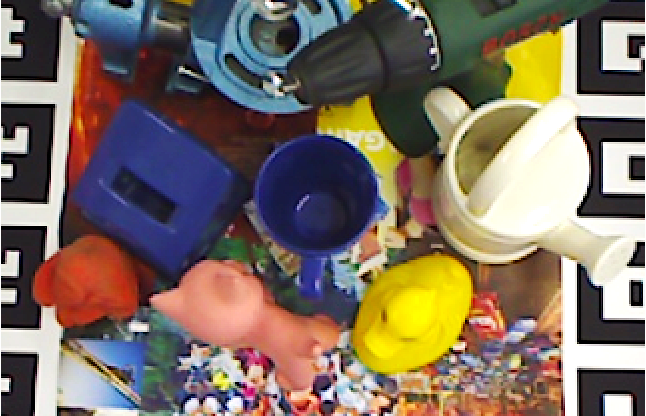}
        \end{minipage}
        \begin{minipage}{0.22\textwidth}
            \includegraphics[width=\textwidth]{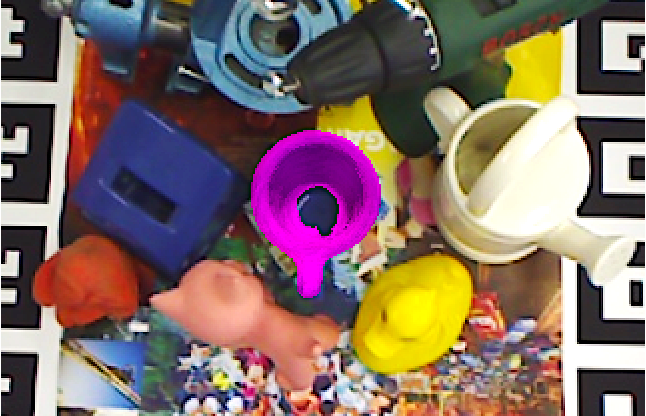}
        \end{minipage}
        \begin{minipage}{0.22\textwidth}
            \includegraphics[width=\textwidth]{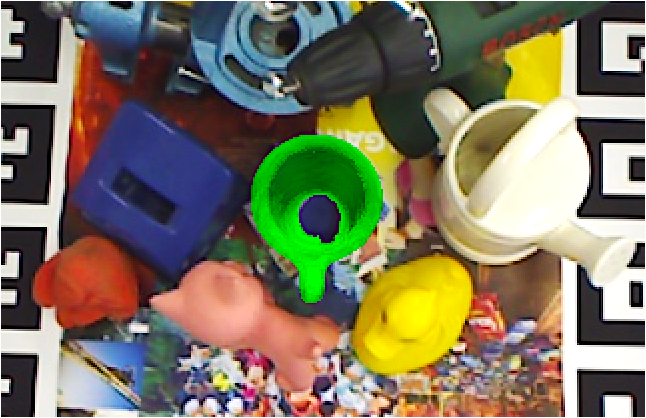}
        \end{minipage}
        \begin{minipage}{0.22\textwidth}
            \includegraphics[width=\textwidth]{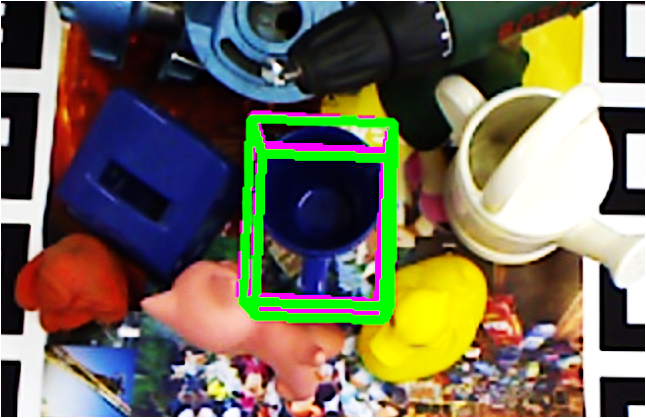}
        \end{minipage}
    \end{minipage}
    \hspace{2em}
    \vspace{1em}
    \begin{minipage}[t]{0.45\textwidth}
        \begin{minipage}[c]{0.05\textwidth}
            \centering
            \rotatebox{90}{Spatula1}
        \end{minipage}
        \begin{minipage}{0.22\textwidth}
            \includegraphics[width=\textwidth]{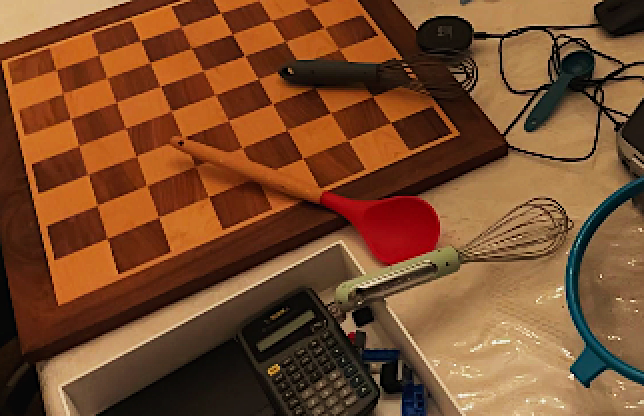}
        \end{minipage}
        \begin{minipage}{0.22\textwidth}
            \includegraphics[width=\textwidth]{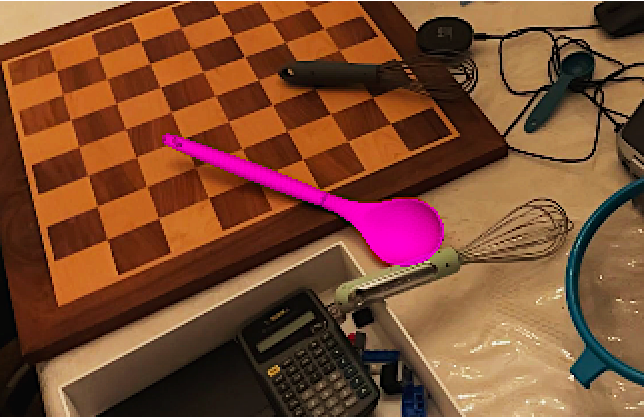}
        \end{minipage}
        \begin{minipage}{0.22\textwidth}
            \includegraphics[width=\textwidth]{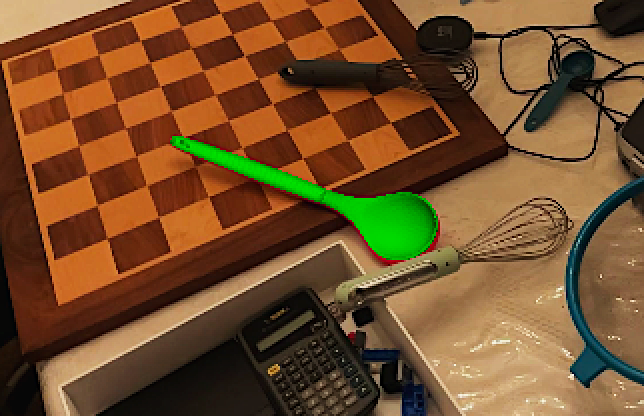}
        \end{minipage}
        \begin{minipage}{0.22\textwidth}
            \includegraphics[width=\textwidth]{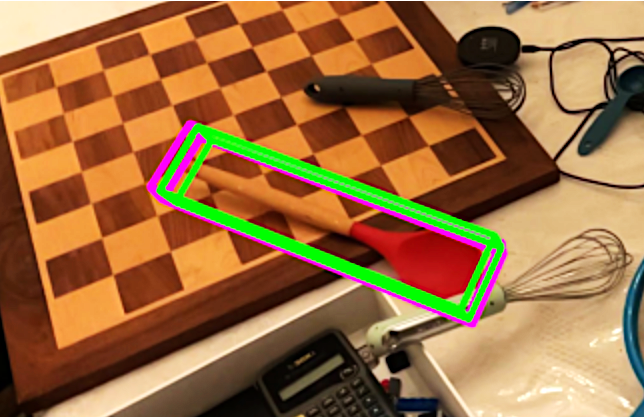}
        \end{minipage}
    \end{minipage}

    \begin{minipage}[t]{0.45\textwidth}
        \begin{minipage}[c]{0.05\textwidth}
            \centering
            \rotatebox{90}{Duck}
        \end{minipage}
        \begin{minipage}{0.22\textwidth}
            \includegraphics[width=\textwidth]{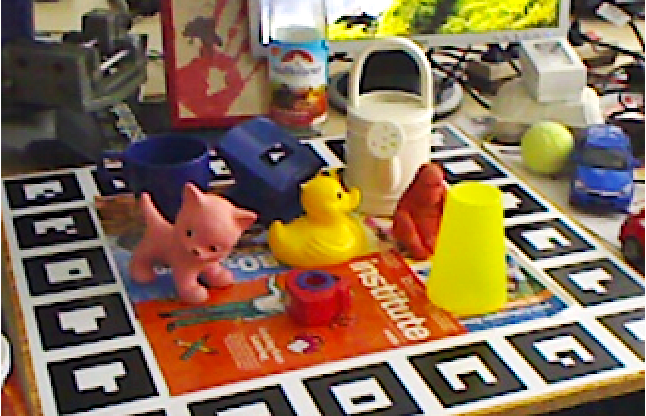}
        \end{minipage}
        \begin{minipage}{0.22\textwidth}
            \includegraphics[width=\textwidth]{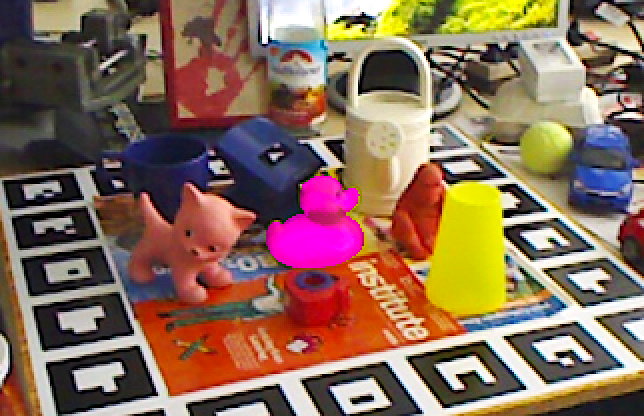}
        \end{minipage}
        \begin{minipage}{0.22\textwidth}
            \includegraphics[width=\textwidth]{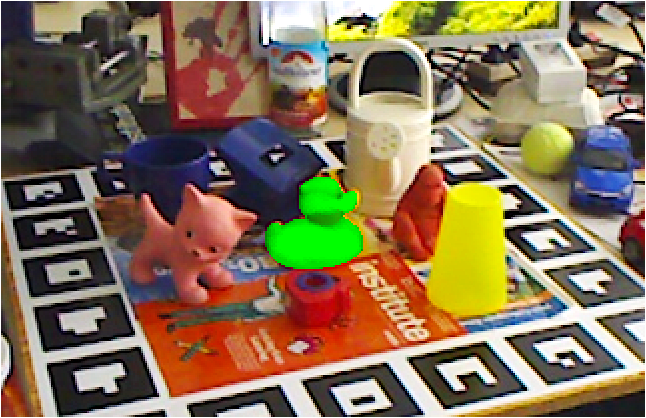}
        \end{minipage}
        \begin{minipage}{0.22\textwidth}
            \includegraphics[width=\textwidth]{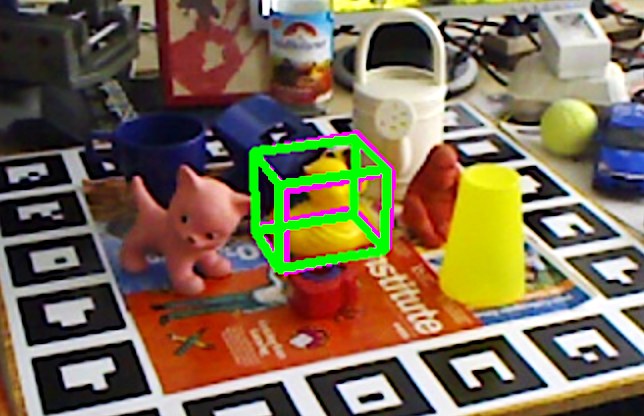}
        \end{minipage}
    \end{minipage}
    \hspace{2em}
    \vspace{1em}
    \begin{minipage}[t]{0.45\textwidth}
        \begin{minipage}[c]{0.05\textwidth}
            \centering
            \rotatebox{90}{Ladle}
        \end{minipage}
        \begin{minipage}{0.22\textwidth}
            \includegraphics[width=\textwidth]{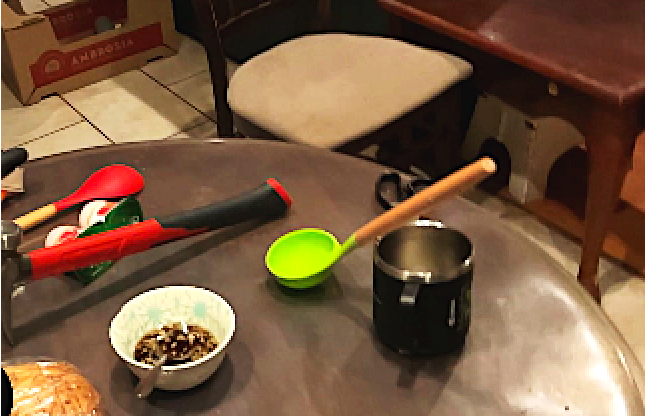}
        \end{minipage}
        \begin{minipage}{0.22\textwidth}
            \includegraphics[width=\textwidth]{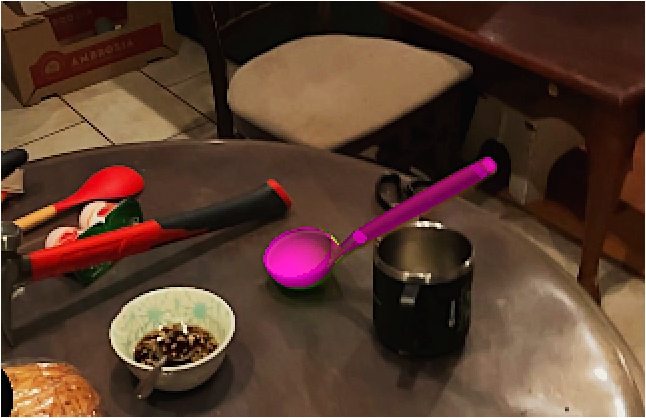}
        \end{minipage}
        \begin{minipage}{0.22\textwidth}
            \includegraphics[width=\textwidth]{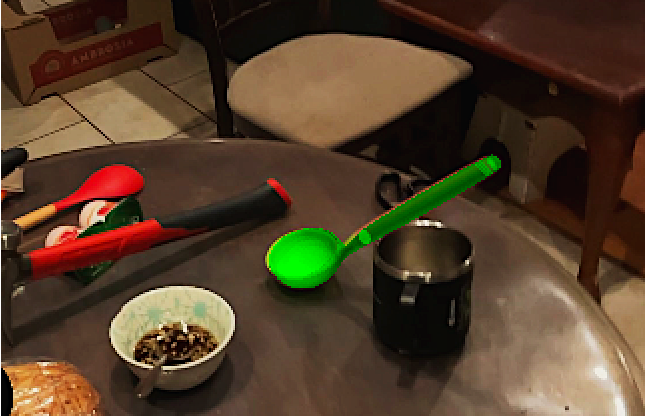}
        \end{minipage}
        \begin{minipage}{0.22\textwidth}
            \includegraphics[width=\textwidth]{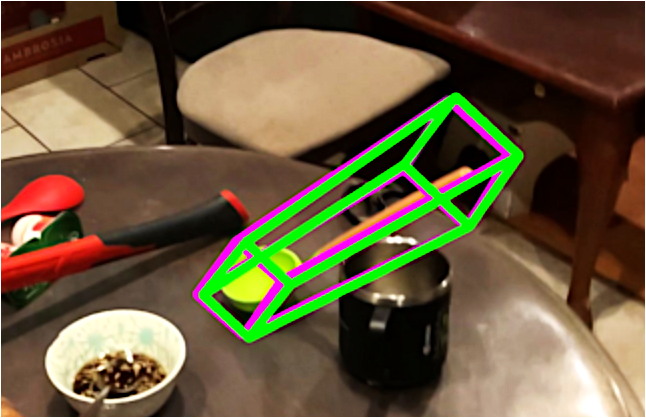}
        \end{minipage}
    \end{minipage}

    \begin{minipage}[t]{0.45\textwidth}
        \begin{minipage}[c]{0.05\textwidth}
            \centering
            \rotatebox{90}{Glue}
        \end{minipage}
        \begin{minipage}{0.22\textwidth}
            \includegraphics[width=\textwidth]{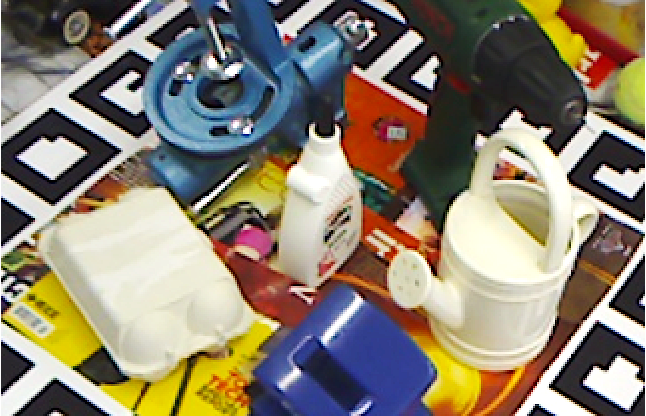}
        \end{minipage}
        \begin{minipage}{0.22\textwidth}
            \includegraphics[width=\textwidth]{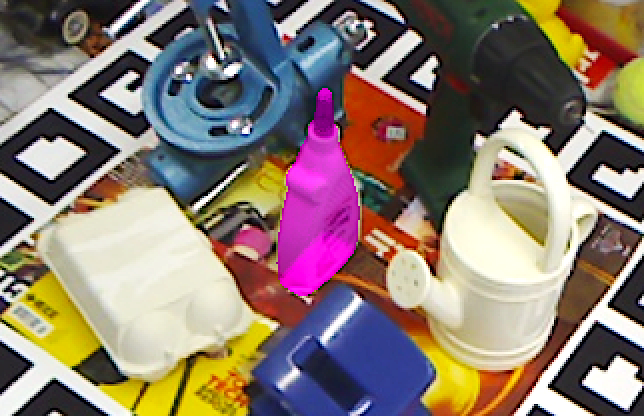}
        \end{minipage}
        \begin{minipage}{0.22\textwidth}
            \includegraphics[width=\textwidth]{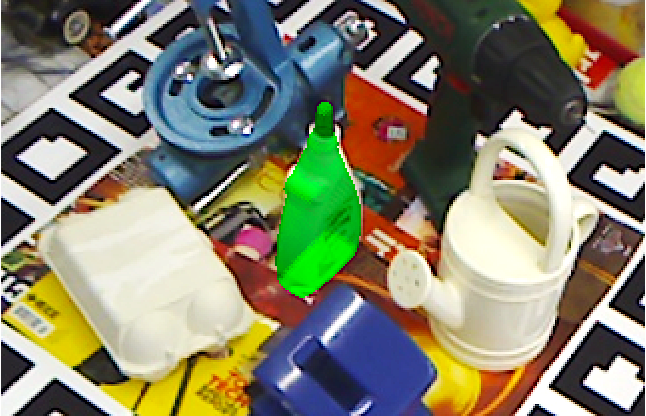}
        \end{minipage}
        \begin{minipage}{0.22\textwidth}
            \includegraphics[width=\textwidth]{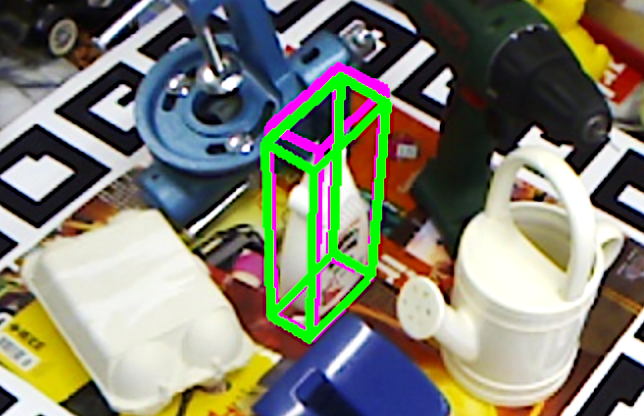}
        \end{minipage}
    \end{minipage}
    \hspace{2em}
    \vspace{1em}
    \begin{minipage}[t]{0.45\textwidth}
        \begin{minipage}[c]{0.05\textwidth}
            \centering
            \rotatebox{90}{Utensil2}
        \end{minipage}
        \begin{minipage}{0.22\textwidth}
            \includegraphics[width=\textwidth]{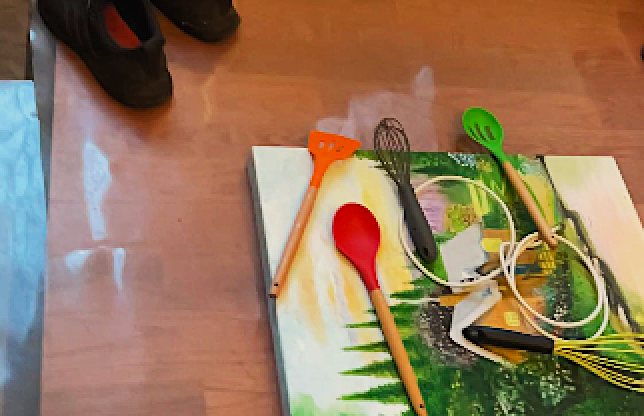}
        \end{minipage}
        \begin{minipage}{0.22\textwidth}
            \includegraphics[width=\textwidth]{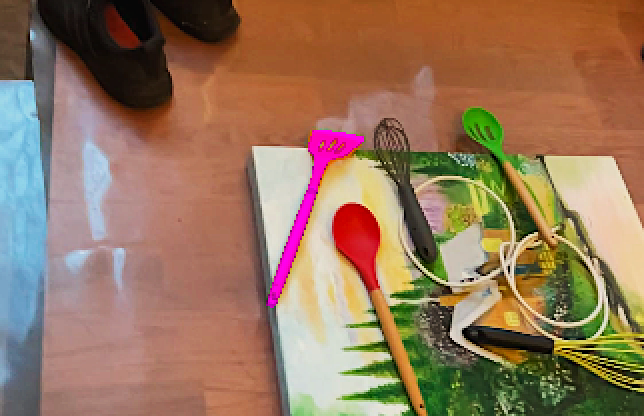}
        \end{minipage}
        \begin{minipage}{0.22\textwidth}
            \includegraphics[width=\textwidth]{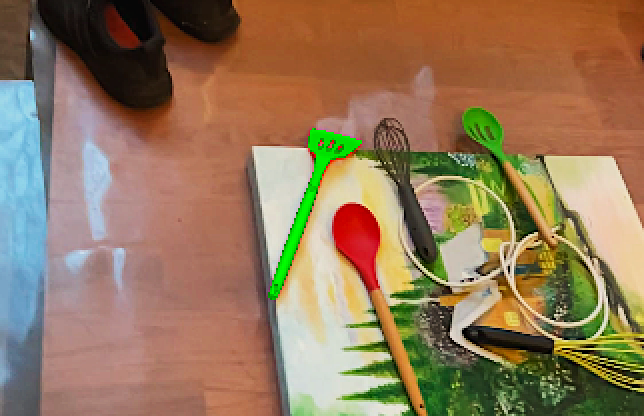}
        \end{minipage}
        \begin{minipage}{0.22\textwidth}
            \includegraphics[width=\textwidth]{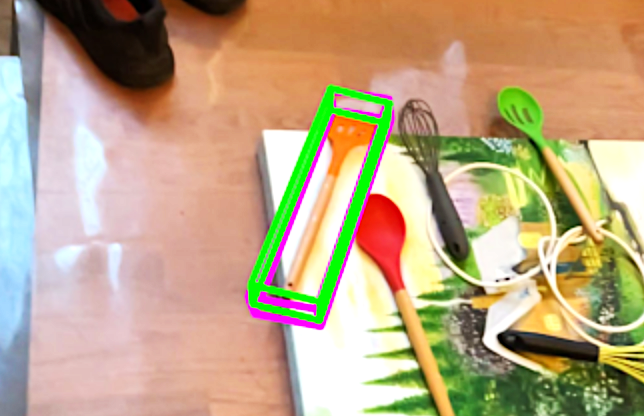}
        \end{minipage}
    \end{minipage}

    \begin{minipage}[t]{0.45\textwidth}
        \begin{minipage}[c]{0.05\textwidth}
            \centering
            \rotatebox{90}{Iron}
        \end{minipage}
        \begin{minipage}{0.22\textwidth}
            \includegraphics[width=\textwidth]{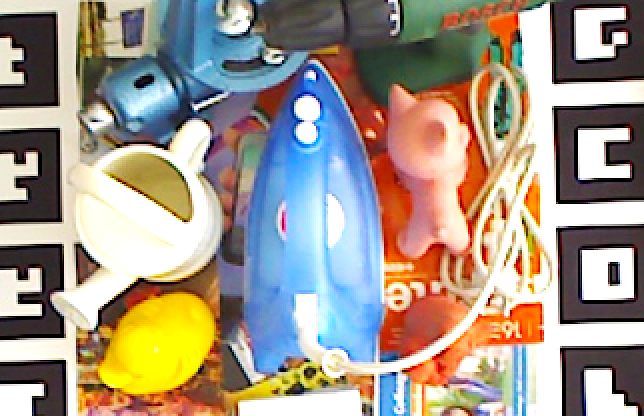}
        \end{minipage}
        \begin{minipage}{0.22\textwidth}
            \includegraphics[width=\textwidth]{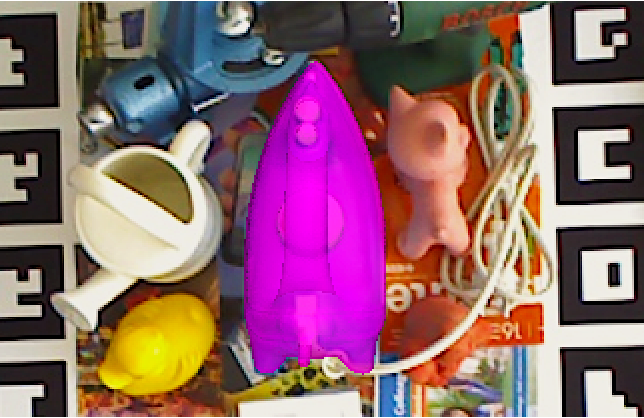}
        \end{minipage}
        \begin{minipage}{0.22\textwidth}
            \includegraphics[width=\textwidth]{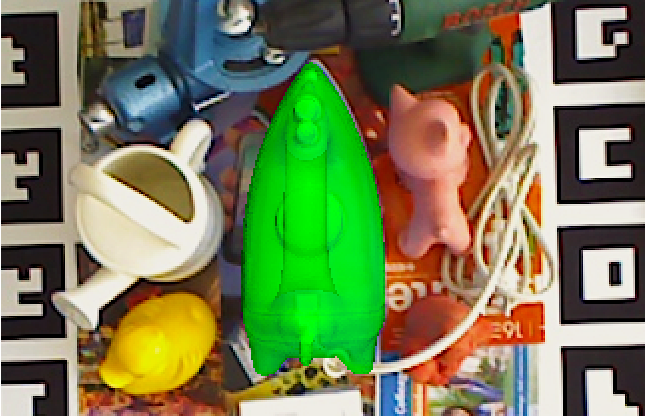}
        \end{minipage}
        \begin{minipage}{0.22\textwidth}
            \includegraphics[width=\textwidth]{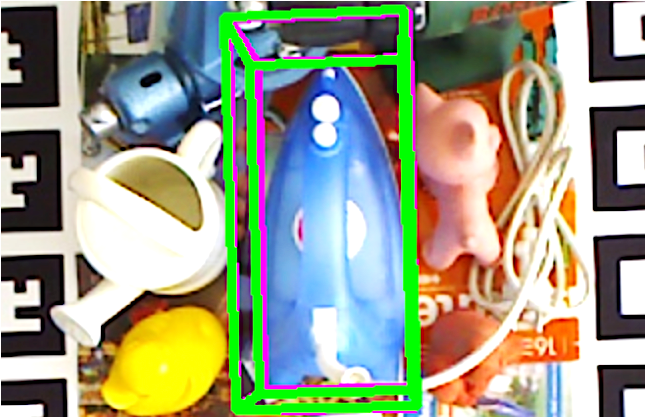}
        \end{minipage}
    \end{minipage}
    \hspace{2em}
    \vspace{1em}
    \begin{minipage}[t]{0.45\textwidth}
        \begin{minipage}[c]{0.05\textwidth}
            \centering
            \rotatebox{90}{Spatula2}
        \end{minipage}
        \begin{minipage}{0.22\textwidth}
            \includegraphics[width=\textwidth]{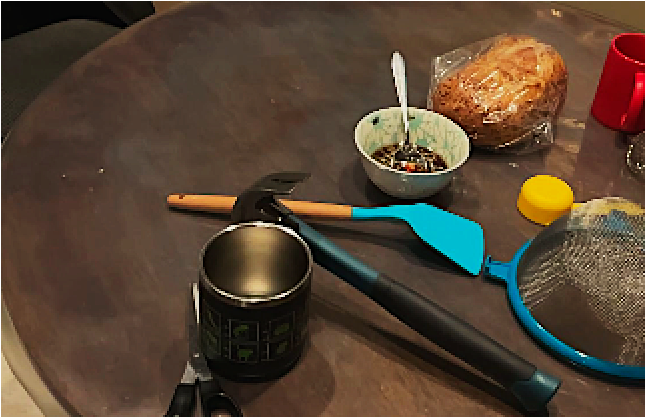}
        \end{minipage}
        \begin{minipage}{0.22\textwidth}
            \includegraphics[width=\textwidth]{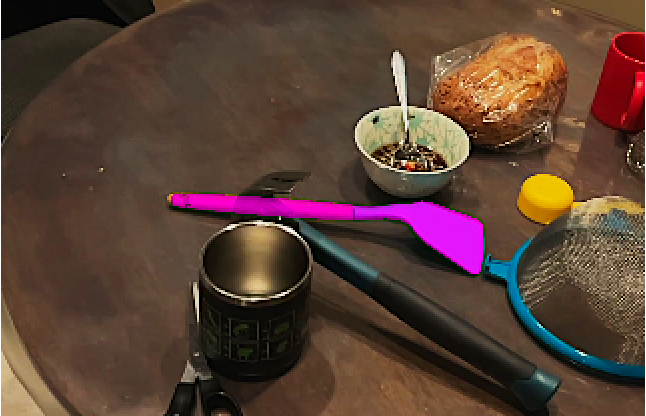}
        \end{minipage}
        \begin{minipage}{0.22\textwidth}
            \includegraphics[width=\textwidth]{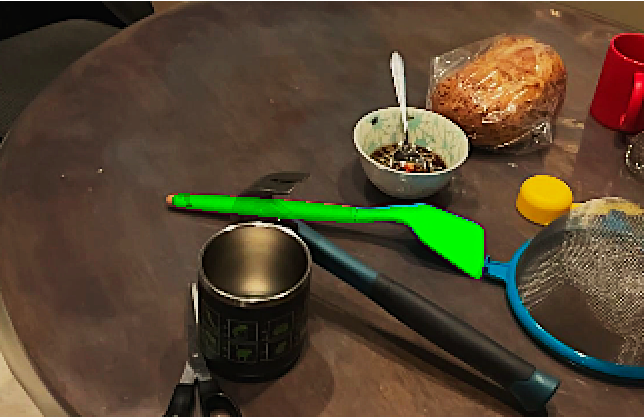}
        \end{minipage}
        \begin{minipage}{0.22\textwidth}
            \includegraphics[width=\textwidth]{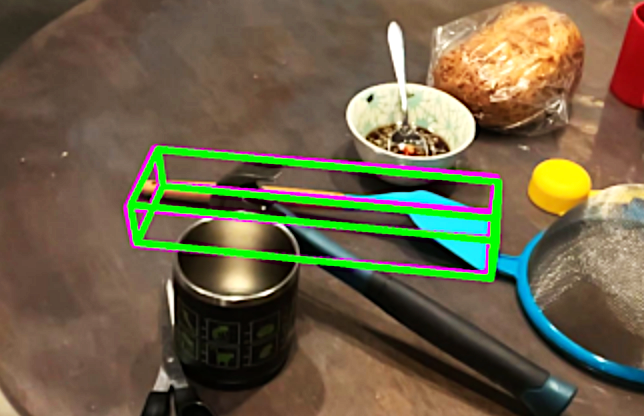}
        \end{minipage}
    \end{minipage}

    \begin{minipage}[t]{0.45\textwidth}
        \begin{minipage}[c]{0.05\textwidth}
            \centering
            \rotatebox{90}{Lamp}
        \end{minipage}
        \begin{minipage}{0.22\textwidth}
            \includegraphics[width=\textwidth]{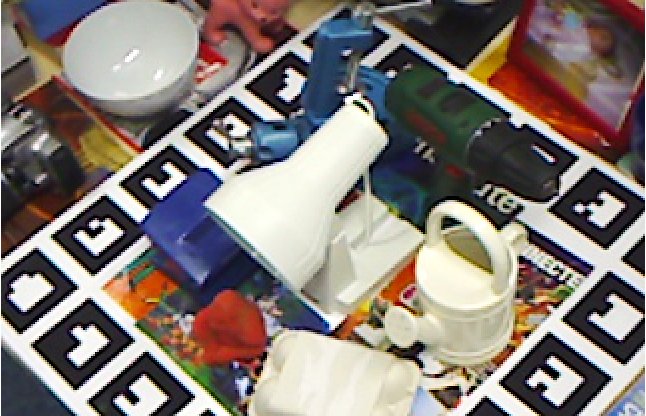}
        \end{minipage}
        \begin{minipage}{0.22\textwidth}
            \includegraphics[width=\textwidth]{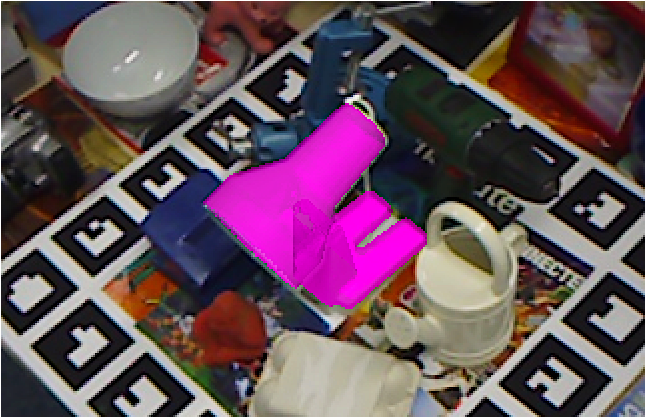}
        \end{minipage}
        \begin{minipage}{0.22\textwidth}
            \includegraphics[width=\textwidth]{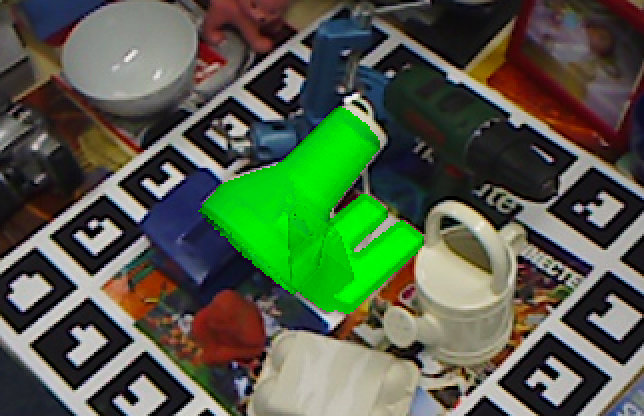}
        \end{minipage}
        \begin{minipage}{0.22\textwidth}
            \includegraphics[width=\textwidth]{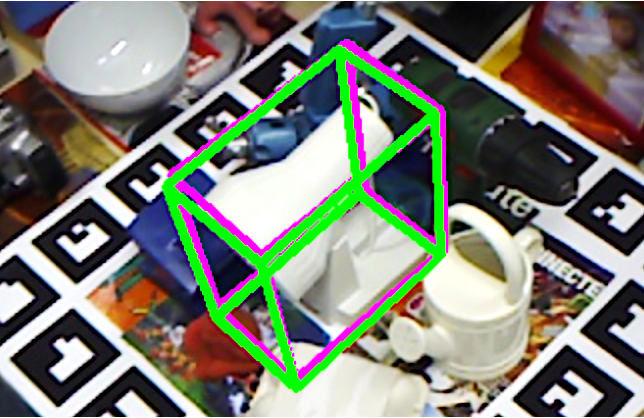}
        \end{minipage}
    \end{minipage}
    \hspace{2em}
    \vspace{1em}
    \begin{minipage}[t]{0.45\textwidth}
        \begin{minipage}[c]{0.05\textwidth}
            \centering
            \rotatebox{90}{Strainer1}
        \end{minipage}
        \begin{minipage}{0.22\textwidth}
            \includegraphics[width=\textwidth]{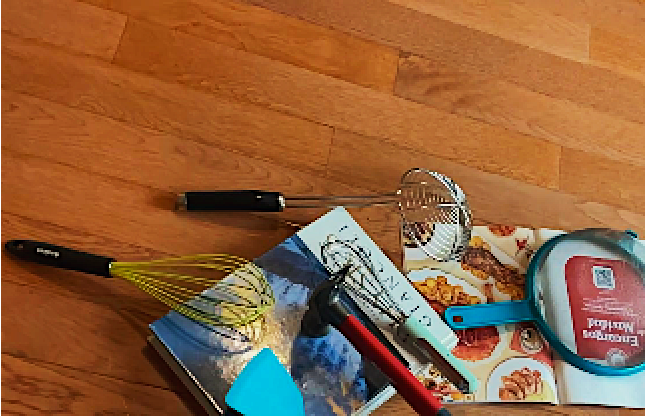}
        \end{minipage}
        \begin{minipage}{0.22\textwidth}
            \includegraphics[width=\textwidth]{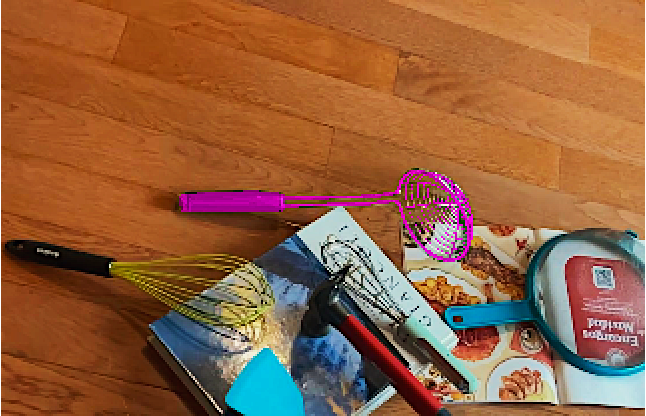}
        \end{minipage}
        \begin{minipage}{0.22\textwidth}
            \includegraphics[width=\textwidth]{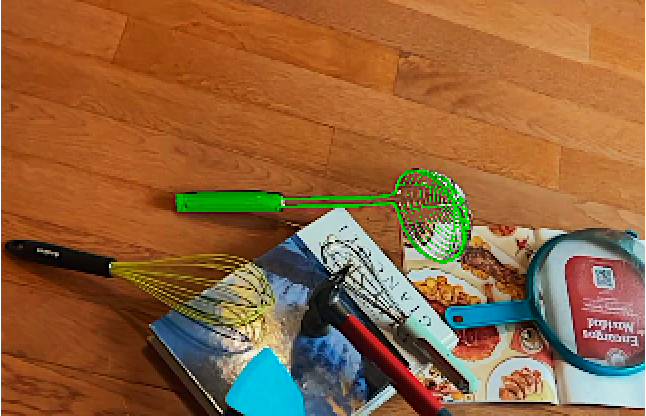}
        \end{minipage}
        \begin{minipage}{0.22\textwidth}
            \includegraphics[width=\textwidth]{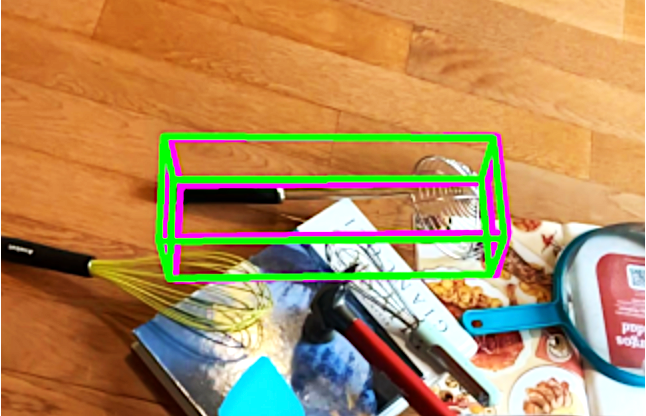}
        \end{minipage}
    \end{minipage}

    \begin{minipage}[t]{0.45\textwidth}
        \begin{minipage}[c]{0.05\textwidth}
            \centering
            \rotatebox{90}{Phone}
        \end{minipage}
        \begin{minipage}{0.22\textwidth}
            \includegraphics[width=\textwidth]{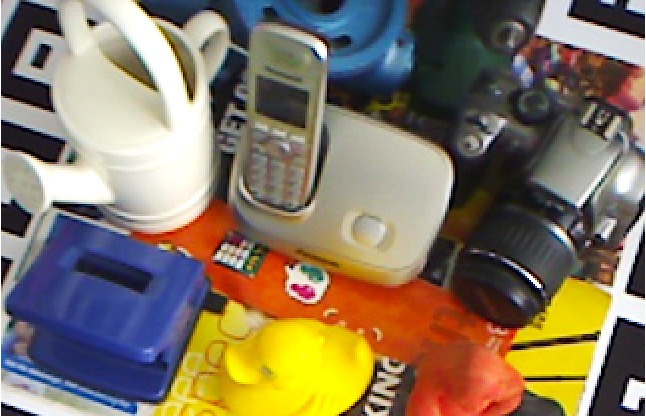}
            \centering
            \footnotesize{Original Image}
        \end{minipage}
        \begin{minipage}{0.22\textwidth}
            \includegraphics[width=\textwidth]{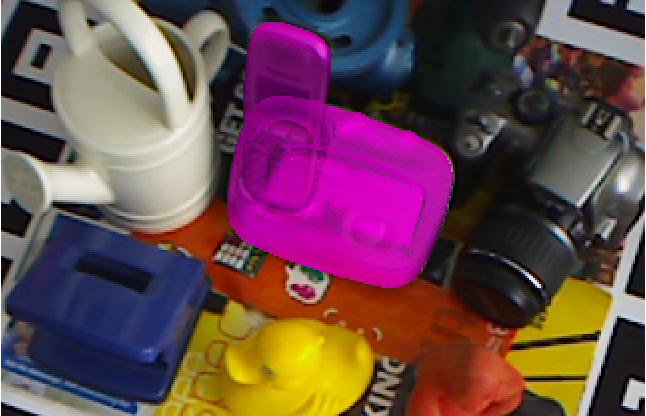}
            \centering
            \footnotesize{Annotated Pose}
        \end{minipage}
        \begin{minipage}{0.22\textwidth}
            \includegraphics[width=\textwidth]{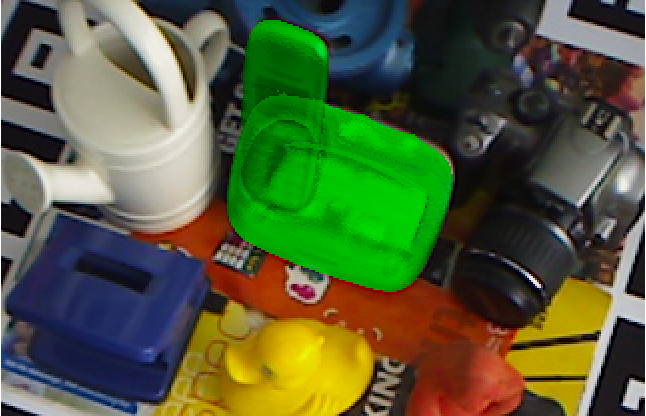}
            \centering
            \footnotesize{GT Pose}
        \end{minipage}
        \begin{minipage}{0.22\textwidth}
            \includegraphics[width=\textwidth]{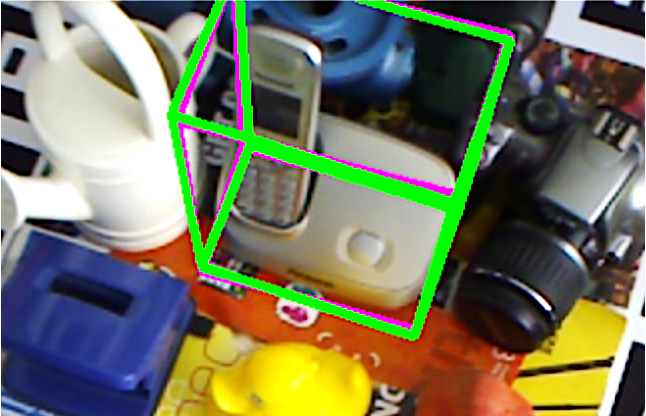}
            \centering
            \footnotesize{Differences}
        \end{minipage}
        
        \centering
        \vspace{1em}Linemod Dataset
    \end{minipage}
    \hspace{2em}
    \vspace{1em}
    \begin{minipage}[t]{0.45\textwidth}
        \begin{minipage}[c]{0.05\textwidth}
            \centering
            \rotatebox{90}{Strainer2}
        \end{minipage}
        \begin{minipage}{0.22\textwidth}
            \includegraphics[width=\textwidth]{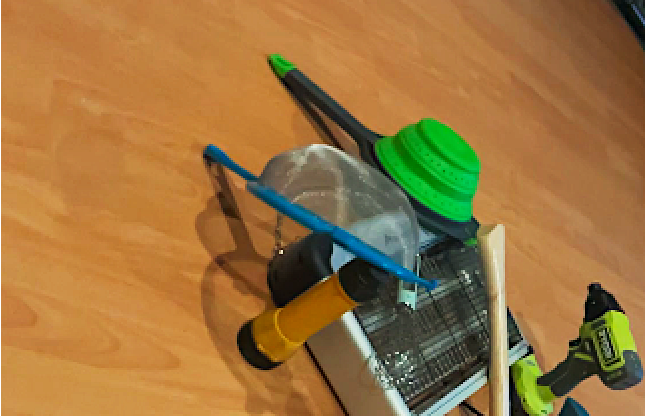}
            \centering
            \footnotesize{Original Image}
        \end{minipage}
        \begin{minipage}{0.22\textwidth}
            \includegraphics[width=\textwidth]{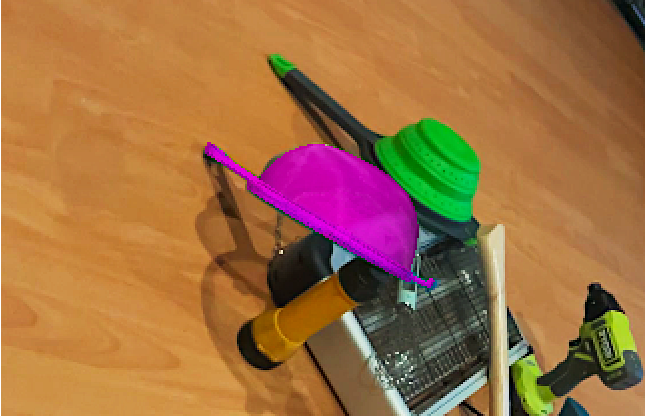}
            \centering
            \footnotesize{Annotated Pose}
        \end{minipage}
        \begin{minipage}{0.22\textwidth}
            \includegraphics[width=\textwidth]{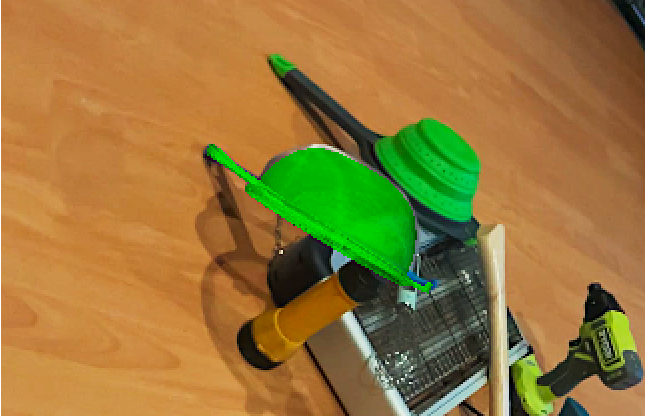}
            \centering
            \footnotesize{GT Pose}
        \end{minipage}
        \begin{minipage}{0.22\textwidth}
            \includegraphics[width=\textwidth]{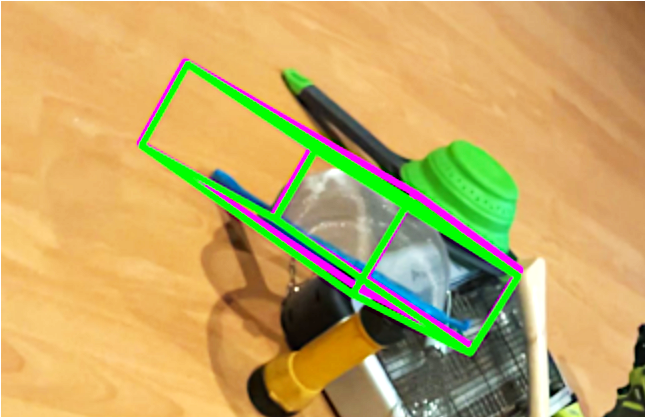}
            \centering
            \footnotesize{Differences}
        \end{minipage}
        
        \centering
        \vspace{1em}HANDAL Dataset
    \end{minipage}
    \caption{\textbf{Qualitative Evaluation of Inter-personal Variability.} This evaluation compares user annotations for 10 objects across two datasets, showing (1) the original image, (2) user-annotated pose, (3) ground-truth pose, and (4) pose differences.}
    \label{fig:inter_personal_qualitative}
\end{figure*}

\subsection{Study Design}

\subsubsection{Participants}
For this study, we recruited 11 participants via a university mailing list. Notably, the participants have a broad spectrum of expertise in 6D pose estimation, ranging from limited understanding to actively working on pose estimation-related projects. This diverse skill set created an ideal experimental environment for assessing the intuitiveness and accessibility of our proposed tool in a real-world setting, where users may have varying levels of domain knowledge.

\subsubsection{Stimuli}
We use two publicly available datasets, HANDAL \cite{guo2023handaldatasetrealworldmanipulable} and Linemod \cite{linemod} for this study. Both of which serve as benchmark datasets for 6D pose estimation. From our perspective, HANDAL is a newer and more challenging dataset than Linemod, presenting greater difficulty for users during annotation. These two datasets provide a comprehensive evaluation of how Vision6D assists users in annotating under varying levels of dataset complexity. We assign five participants to annotate on the Linemod dataset, and six participants to annotate on the HANDAL dataset. 

From each dataset, we selected 10 distinct samples, with each sample representing a unique 6D pose scenario that participants were required to annotate. The selection ensures a diverse range of poses and object configurations, providing a robust basis for evaluating the accuracy and consistency of annotations generated using Vision6D. To thoroughly evaluate both intra-personal (consistency within the same individual) and inter-personal (variability between different individuals) pose annotation performance, each of these 10 samples is shuffled and presented three times during the pose annotation session. This approach ensures multiple annotations per sample from each participant. More specifically, every participant is tasked with annotating each sample three separate times, with the order of the samples randomized for each iteration.

\subsubsection{Protocol}
Participants in the experiment were required to complete a series of tasks designed to assess both the usability and precision of Vision6D. Initially, each participant was informed about the basic commands and shortcuts related to the user interface and functionalities. During the introduction session, some essential functions, such as 2D and 3D object manipulation and the annotation process were explained, ensuring that all participants understood how to use the system.

In the pose annotation session, participants were tasked with annotating 6D poses for a set of 3D objects projected onto 2D scenes. For each annotation task, they were provided with a 2D image where the corresponding 3D object was placed in some random place in the scene, and they were asked to adjust the object's orientation and position using the interactive controls until they confirmed the 3D object pose accurately matched the object's placement in the 2D image scene. The tasks varied in complexity, with some scenarios presenting additional challenges such as partial occlusions to better simulate real-world environments.

\subsubsection{Data Collection}
Throughout the experiment, our tool recorded quantitative metrics such as the time taken to complete each annotation and the pose deviation between the participant’s annotations and the ground-truth data. Additionally, after completing the annotation tasks, participants were asked to fill out a set of questionnaires to gather feedback on the tool’s intuitiveness, ease of use, and overall effectiveness via the standardized NASA-TLX \cite{HART1988139} and SUS surveys \cite{sus}. This combined approach ensured that both performance data and user experiences were captured, providing a holistic evaluation of Vision6D. For the following experiments, we present the top 90\% results to eliminate outliers and reduce the impact of extreme deviations. 

\section{Results}
We evaluated the accuracy of 6D pose annotations using the intra-personal and inter-personal variability metrics. Intra-personal consistency quantifies how an individual reproduces the same pose annotation across multiple trials, providing insights into the tool’s robustness and reliability. Inter-personal variability, on the other hand, measures the differences in pose error among different participants, capturing variations in human perception and interaction with the proposed 3D user interface. By analyzing these metrics, we aim to determine the pose reproducibility and annotation reliability when using Vision6D, as well as the overall robustness and effectiveness of the tool in generating accurate 6D pose annotations across a diverse range of users with varying levels of expertise.
\subsection{Annotation Accuracy}
To assess the accuracy of user-annotated 6D poses, we evaluate both inter-personal variability across different users and intra-personal consistency. These results help us to understand the precision and robustness of using Vision6D in various pose annotation scenarios.

\subsubsection{Inter-personal Variability}
To evaluate inter-personal variability, we compare the pose annotation errors against the ground-truth pose across different participants for the same sample, as shown in Figure~\ref{fig:inter_personal_consistency}. S0 to S9 represent different samples for both Linemod and HANDAL datasets. We quantify the error variation by calculating the mean Euclidean distance (in mm), angular distance (in degrees), and the Average Distance Metric (ADD) (in mm). These results are calculated using each participant’s best pose annotation from the three repeated trials, defined as the one with the lowest ADD score, and compared to the corresponding object ground-truth poses from each dataset. 

A total of 11 participants contributed to this analysis for annotating two datasets. The angular distance metric is defined in Equation~\ref{eq:d_rot}, which measures the rotational difference between two rotation matrices, $R$ and $R'$. Additionally, the ADD metric quantifies the discrepancy between two 3D object poses by computing the mean Euclidean distance between their corresponding model points. For a set of model points $\{v_i\}^{N}_{i=1}$ and two poses $(R_1, t_1)$ and $(R_2, t_2)$, the ADD is computed following Equation~\ref{eq:add}.
\begin{equation}
    D_{rot} = \arccos\left(\frac{\mathrm{Tr}(R^T R') - 1}{2}\right)
\label{eq:d_rot}
\end{equation}
\begin{equation}
    D_{add} = \frac{1}{N}\sum_{i=1}^{N} \lVert (R_1v_i+t_1) - (R_2v_i + t_2) \lVert
\label{eq:add}
\end{equation}
From the figure, the mean angular distance error is $4.77^\circ \pm 2.84$ for the Linemod dataset and $5.88^\circ \pm 4.68$ for the HANDAL dataset. The mean Euclidean distance error is 14.91 $\pm$ 9.43 mm for Linemod and 29.27 $\pm$ 25.74 mm for HANDAL. Similarly, the mean ADD error is 15.27 $\pm$ 9.33 mm for Linemod and 30.53 $\pm$ 26.72 mm for HANDAL.

The inter-personal angular distance, Euclidean distance, and ADD errors remain relatively low across most samples, indicating that users consistently annotate object positions with high accuracy. However, the distribution differences between the two datasets reveal that certain objects have larger annotation errors, such as S7 in the Linemod dataset and S5 in the HANDAL dataset. This is likely due to increased object complexity or occlusion challenges commonly found in the HANDAL dataset. The results highlight the effectiveness of Vision6D in supporting accurate annotations across different datasets. However, the increased complexity of the HANDAL dataset introduces greater annotation variability, aligning with our observation that it presents more challenges for users.

To further demonstrate the participants' annotations across different samples, Figure~\ref{fig:inter_personal_qualitative} presents a qualitative evaluation, and the results suggest that Vision6D successfully assists users to accurately align and register 3D objects to the 2D scene by providing essential visual cues.
\begin{table*}[!ht]
    \centering
    \begin{tabular}{l*{6}{c}}
        \toprule
        \multirow{2}{*}{Dataset} & \multicolumn{6}{c}{Mean Annotation Time per Person [s]} \\
        \cmidrule(lr){2-7}
         & L0/H0 & L1/H1 & L2/H2 & L3/H3 & L4/H4 & L5/H5 \\
        \midrule
        Linemod & 148.5 & 121.7 & 54.34 & 101.1 & 85.29 & -- \\
        HANDAL  & 56.73 & 99.32 & 154.7 & 105.4 & 94.84 & 71.70 \\
        \bottomrule
    \end{tabular}
    \caption{\textbf{Mean Annotation Time per Person.} Mean annotation time (in seconds) is reported for 11 participants across two datasets (Linemod and HANDAL) and six conditions (L0/H0 to L5/H5).}
    \label{Tab:mean_annotation_time_per_person}
\end{table*}
\subsubsection{Intra-personal Consistency}
To assess intra-personal consistency, each participant annotated the same sample three times, with the order randomized to minimize potential bias. We perform pairwise comparisons between the three trials while using the same evaluation metrics as in inter-personal variability analysis, such as Euclidean distance, angular distance, and ADD, to comprehensively measure annotation consistency among the same users.
The intra-personal consistency score for each participant is calculated by averaging these pairwise metrics across multiple trials over 10 samples. Lower Euclidean distance, ADD score, and angular distance indicate that Vision6D allows users to consistently replicate their annotations with minimal variance.

The quantitative analysis, shown in Figure~\ref{fig:intra_personal_consistency}, demonstrates the robustness of our tool in generating consistent 6D pose annotations across multiple trials. L0 to L4 represent different users who annotated the Linemod dataset, while H0 to H5 correspond to users who annotated the HANDAL dataset. From the figure, the mean angular distance error is $7.11^\circ$ for the Linemod dataset and $10.08^\circ$ for the HANDAL dataset. The mean Euclidean distance error is 19.17 mm for Linemod and 32.34 mm for HANDAL. Similarly, the mean ADD error is 21.28 mm for Linemod and 40.46 mm for HANDAL. The intra-personal average angular distance remains relatively low for most users in both datasets, indicating that Vision6D allows users to annotate poses with minimal rotational deviation among trials. The Euclidean distance has higher variance among users, particularly in the HANDAL dataset, showing the increased complexity inherent in this dataset. These results align with our observation on the two datasets. While there are variations among participants, the average ADD metric remains within a reasonable range of 50 mm. The outcomes highlight the robustness and reproducibility of Vision6D by showing that participants consistently annotate 6D poses with low intra-personal variance.

\subsection{Annotation Efficiency}
To evaluate the efficiency of the Vision6D annotation process, we measure the time taken (in seconds) by each participant to annotate a single sample. This metric provides insights into how intuitive and user-friendly the tool is, as well as how quickly users can generate accurate 6D pose annotations. For each participant, we record the annotation time for all samples across the two datasets, capturing the time required to complete each annotation task. We then analyze time-related metrics to evaluate efficiency across participants.
To evaluate the annotation efficiency, we present Table~\ref{Tab:mean_annotation_time_per_person}, which summarizes the average annotation time recorded across 11 participants for the two datasets. As shown in the table, the Linemod dataset required an average of 102.19 $\pm$ 31.93 seconds per annotation, while the HANDAL dataset required 97.11 $\pm$ 30.74 seconds. The annotation time for the Linemod dataset has a wider range across participants, with time ranging from 54.34 seconds (L2) to 148.5 seconds (L0). This suggests that some users found the samples easy to annotate, while others required more time to complete. Similarly, the HANDAL dataset also showed wide variation, with annotation time ranging from 56.73 seconds (H0) to 154.7 seconds (H2). These results demonstrate that Vision6D provides fast 6D pose annotation across datasets, maintaining short annotation time for both Linemod and HANDAL, despite the latter being a more complex dataset. These findings indicate that Vision6D provides an intuitive interface that enables efficient pose annotations for users with varying levels of expertise.
\begin{figure*}[ht]
    \centering
    \begin{minipage}[t]{0.4\textwidth}
        \begin{minipage}[c]{0.05\textwidth}
            \centering
            \rotatebox{90}{NASA-TLX Metric}
        \end{minipage}
        \begin{minipage}{0.95\textwidth}
            \includegraphics[width=\textwidth]{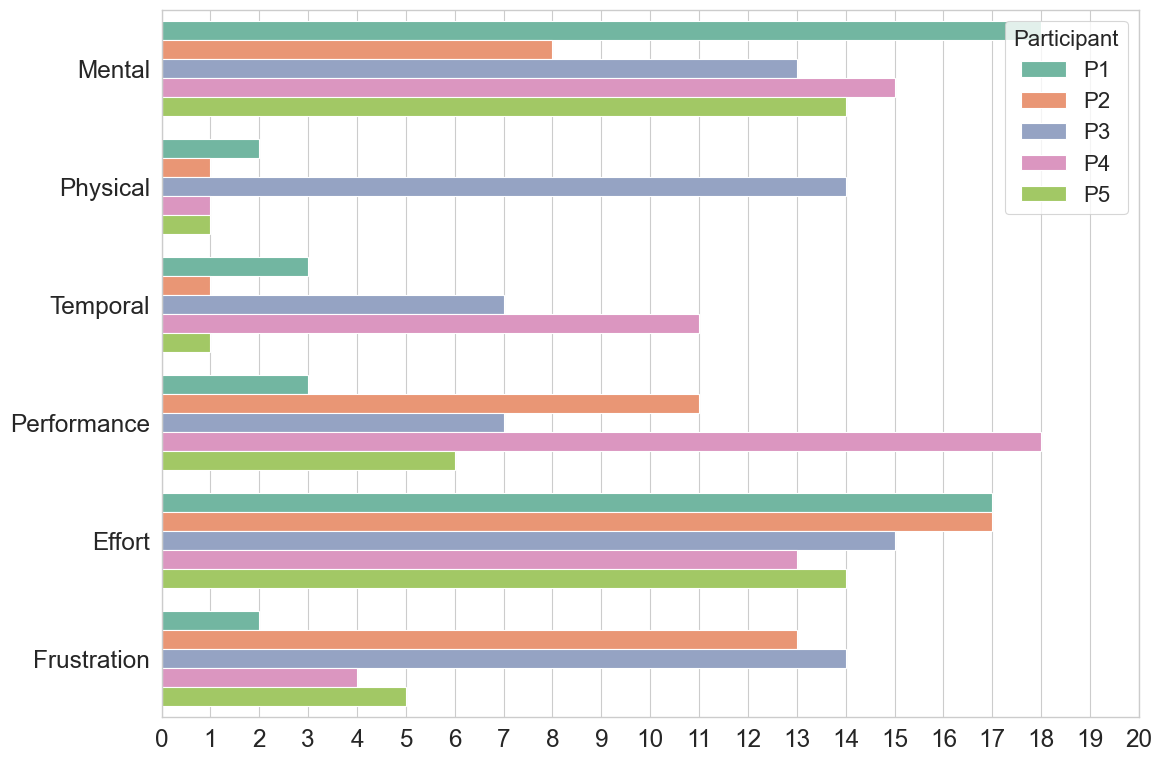}
            \centering
            Metric Scores
        \end{minipage}\hspace{-0.3em}
    \end{minipage}
    \hspace{3em}
    \begin{minipage}[t]{0.4\textwidth}
        \begin{minipage}[c]{0.05\textwidth}
            \centering
            \rotatebox{90}{NASA-TLX Metric}
        \end{minipage}
        \begin{minipage}{0.95\textwidth}
            \includegraphics[width=\textwidth]{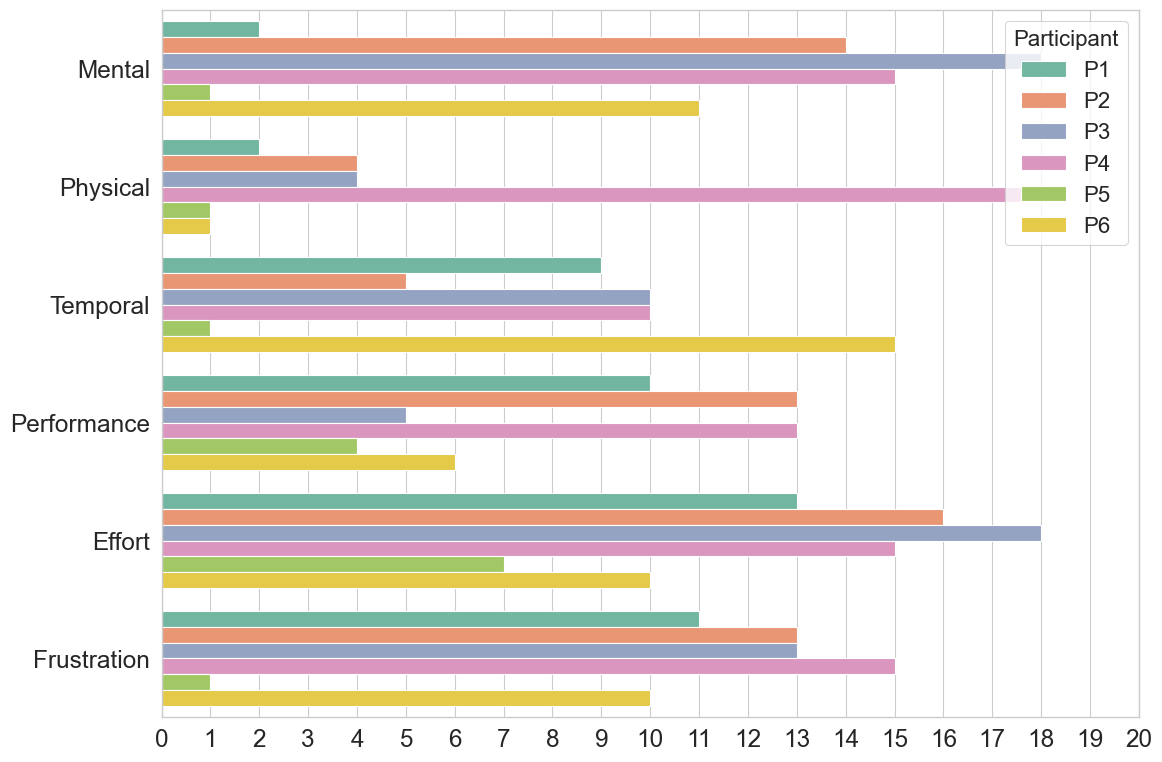}
            \centering
            Metric Scores
        \end{minipage}\hspace{-0.3em}
    \end{minipage}
    \vspace{1em}
    
    \begin{minipage}[t]{0.4\textwidth}
        \begin{minipage}[c]{0.05\textwidth}
            \centering
            \rotatebox{90}{SUS Questions}
        \end{minipage}
        \begin{minipage}{0.95\textwidth}
            \includegraphics[width=\textwidth]{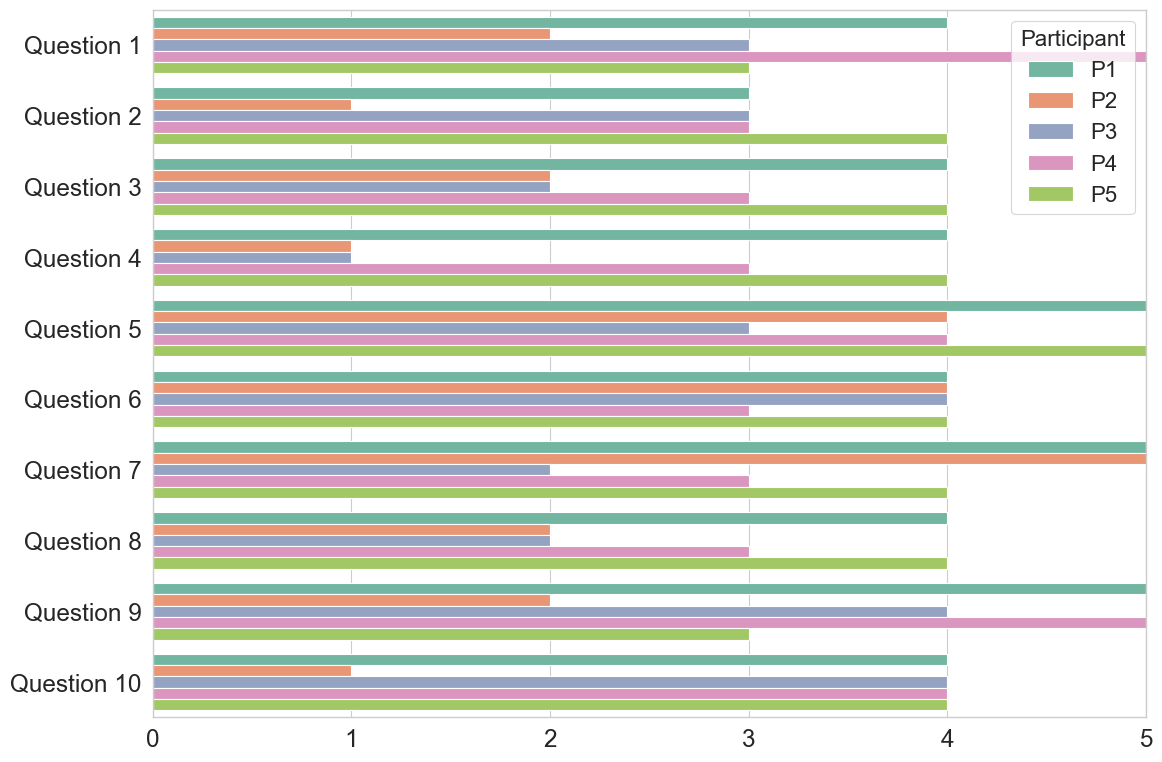}
            \centering
            Adjusted Scores
        \end{minipage}\hspace{-0.3em}

        \centering
        Linemod Dataset
    \end{minipage}
    \hspace{3em}
    \begin{minipage}[t]{0.4\textwidth}
        \begin{minipage}[c]{0.05\textwidth}
            \centering
            \rotatebox{90}{SUS Questions}
        \end{minipage}
        \begin{minipage}{0.95\textwidth}
            \includegraphics[width=\textwidth]{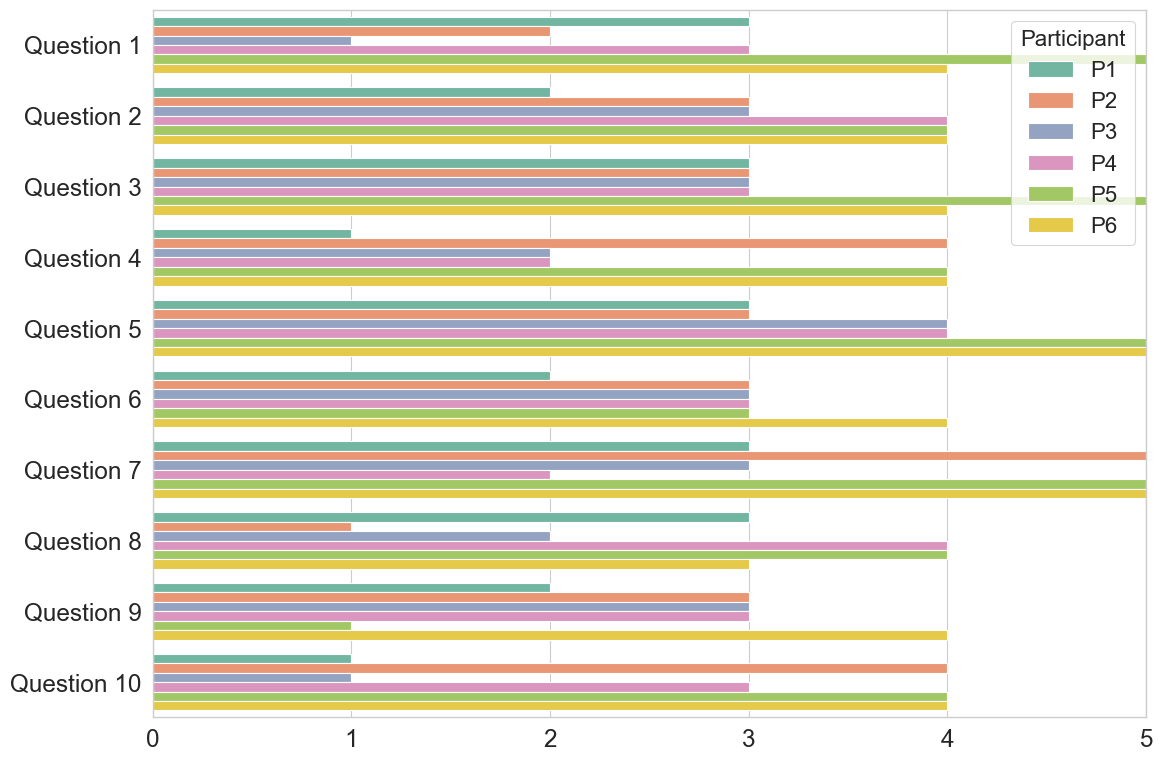}
            \centering
            Adjusted Scores
        \end{minipage}\hspace{-0.3em}
    
        \centering
        HANDAL Dataset
    \end{minipage}

    \caption{\textbf{User Feedback.} We demonstrate the user feedback using two standardized evaluation metrics: NASA-TLX for assessing cognitive workload and task difficulty, and SUS for measuring system usability.}
    \label{fig:user_feedback}
\end{figure*}

\subsection{User Feedback}
To assess the cognitive load and usability associated with Vision6D, we collected user feedback using two standard evaluation instruments: the NASA Task Load Index (NASA-TLX) and the System Usability Scale (SUS). NASA-TLX measures perceived workload across six dimensions: \textit{mental demand}, \textit{physical demand}, \textit{temporal demand}, \textit{performance}, \textit{effort}, and \textit{frustration}, with scores ranging from 0 to 20. Higher NASA-TLX scores indicate higher task difficulty and cognitive effort. SUS provides a standardized measure of system usability, with scores ranging from 1 to 5. It consists of ten standardized questions that assess system usability, alternating between positive and negative statements to minimize response bias. These questions evaluate aspects such as \textit{use frequency} (Question 1), \textit{complexity} (Question 2), \textit{ease of use} (Question 3), \textit{need for technical support or prior learning} (Question 4), \textit{integration of functions} (Question 5), \textit{system consistency} (Question 6), \textit{ease of learn} (Question 7), \textit{system cumbersome} (Question 8), \textit{confidence in use} (Question 9), and \textit{learning effort} (Question 10). We adjust the negative statements through inversion to make all scores relate to a positive statement. Higher adjusted scores indicate a more user-friendly experience, while lower scores suggest usability challenges. Together, these metrics offer valuable insights into the subjective demands of operating the system and its ease of use.
Figure~\ref{fig:user_feedback} presents NASA-TLX and SUS scores across all participants for two datasets: Linemod and HANDAL, offering a detailed view of overall user experiences. Each bar represents the score reported by a participant, with different colors indicating different participants. From the presented figures, the scores vary across participants, reflecting subjective experiences and diverse user profiles.

\subsubsection{NASA Task Load Index (NASA-TLX) Evaluation}
In this evaluation, our findings are summarized in the following for each metric in NASA-TLX.
For the \textit{Mental Demand}, participants reported higher scores for the HANDAL dataset than for Linemod, suggesting that it required more cognitive effort. A higher mental demand score indicates that users had to concentrate to complete tasks, likely due to the data complexity observed in HANDAL.
For the \textit{Physical Demand}, scores were generally low across all participants, meaning that Vision6D does not require excessive manual effort. This suggests that users did not experience significant physical strain while interacting with the system.
For the \textit{Temporal Demand}, scores varied among participants, indicating that some users felt time pressure while others did not.
For the \textit{Performance}, most participants reported high scores, meaning they felt successful in completing their tasks. This suggests that, despite variations in workload, users generally believed they were able to achieve accurate and satisfactory results.
For the \textit{Effort}, scores ranged from moderate to high, indicating that users had to work hard to complete tasks effectively. Higher \textit{Effort} scores suggest that task complexity or system interactions required significant focus and manual adjustments, particularly for more challenging datasets like HANDAL.
For the \textit{Frustration}, scores showed variation across participants, meaning that some users had a smooth experience while others encountered difficulties. Higher \textit{Frustration} scores likely correlate with increased task complexity or annotation challenges, making the experience more demanding for certain users.

\subsubsection{System Usability Scale (SUS) Evaluation}
In this evaluation, most participants consistently gave high scores across all questions, indicating positive feedback on usability. 
For the \textit{use frequency}, most participants rated this question highly, suggesting that Vision6D has strong potential for continued use. 
For the \textit{complexity (inverted)}, the \textit{easy to use}, and the \textit{easy to learn}, scores were consistently high across participants, indicating that most users found the system intuitive and relatively easy to learn.
For the \textit{need for technical support or prior learning (inverted)}, scores were also high, meaning that most participants felt confident using the system independently without requiring external assistance.
For the \textit{integration of functions} and \textit{system consistency}, participants rated Vision6D favorably, suggesting that its tools work well together and provide a seamless experience.
For the \textit{system cumbersome (inverted)}, responses were diverse, meaning that while some users found Vision6D easy to use, others struggled with certain interaction processes, which may have led to frustration in workflow.
For the \textit{confidence in use} and \textit{learning effort (inverted)}, the scores suggest that while some users adapted quickly, others required more time to become comfortable with the system.

Overall, the evaluation results highlight differences in user experiences, indicating that some users found tasks more challenging than others, leading to diverse workload and usability feedback. Our questionnaires suggest that Vision6D's user interface is easy to use and learn, while 6D pose annotations are inherently challenging and require decent effort.

\section{Limitations and Future Work}
One of the inherent limitation in 6D pose annotation and using Vision6D is dealing with round and symmetric texture-less objects, such as spherical objects. These objects often lack distinctive features or asymmetrical markers, making it difficult to determine their exact orientation. Ambiguities in rotation can lead to multiple plausible pose solutions, particularly when the object appears visually identical from different perspectives. For example, a perfectly spherical object has infinite valid rotations about its center, and a cylindrical or radially symmetric object may have rotational ambiguities around its primary axis.

For future work, integrating deep-learning-based pose estimation models or 2D-to-3D PnP registration techniques could significantly accelerate the annotation process by generating initial pose predictions, thereby reducing manual annotation effort.
Furthermore, rather than annotating object poses in individual images, we envision a hybrid approach where Vision6D assists in automatic pose propagation throughout a video sequence. By leveraging the initial annotated pose, camera localization techniques, and geometric tracking, the system can predict and adjust subsequent frames’ object poses. Users will then only need to refine the automatically generated poses instead of starting from scratch for each frame. This approach could substantially accelerate pose annotation workflows, improving Vision6D's efficiency and scalability for large-scale datasets.
Ultimately, our goal is to establish Vision6D as a versatile and widely accessible open-source software for both 6D pose estimation research and real-world applications.

\section{Conclusion}
In this paper, we introduced Vision6D, an interactive 3D-to-2D visualization and pose annotation tool designed to support research in the field of 6D pose estimation. Vision6D is particularly valuable in scenarios where camera pose data are either unavailable or cannot be retrieved from prerecorded videos. By providing precise manual pose annotation to generate 6D camera poses, Vision6D has supported several deep-learning-based studies, including those proposed in \cite{10.1007/978-3-031-64299-9_1, 10.1117/12.3008830, zhang2025ssddgansinglestepdenoisingdiffusion,zhang2024mastoidectomymultiviewsynthesissingle, Davalos2024}. Its contributions have significantly advanced research in diverse domains, such as in education and healthcare settings, proving its value as a powerful tool for retrieving 6D camera poses with minimal requirements.

Furthermore, Vision6D is developed as a cross-platform desktop application, ensuring accessibility and ease of use for researchers and practitioners. The tool allows users to interactively manipulate and annotate 3D objects in 2D scene scenarios, bridging the gap between 2D image observations and 3D geometric placement. Through a comprehensive user study, we have demonstrated that Vision6D enables users to produce accurate and efficient pose annotations. By addressing the challenges above and expanding its capabilities, Vision6D has the potential to become a versatile and widely-used tool for 6D pose annotation and research.

\bibliographystyle{IEEEtran}
\bibliography{references}

\end{document}